\documentclass{aa}  
\usepackage{graphicx}
\usepackage{placeins}
\usepackage{float}
\usepackage{dblfloatfix}
\usepackage{txfonts}
\begin{document}

   \title{Backstreaming ions at a high Mach number interplanetary shock}
    \titlerunning{Dynamics of backstreaming ions at a complex interplanetary shock}
    \authorrunning{A. P. Dimmock et al.,}
   \subtitle{Solar Orbiter measurements during the nominal mission phase}

   \author{A. P. Dimmock
          \inst{1}
          \and
          M. Gedalin\inst{2}
          \and
          A. Lalti\inst{1,3}
          \and
          D. Trotta\inst{4}
          \and
          Yu. V. Khotyaintsev\inst{1}
          \and
          D. B. Graham\inst{1}
          \and
          A. Johlander\inst{5,1}
          \and
          R. Vainio\inst{6}
          \and
          X. Blanco-Cano\inst{7}
           \and
          P. Kajdič\inst{7}
          \and
          C. J. Owen\inst{8}
          \and
          R. F. Wimmer-Schweingruber \inst{9}
          }

   \institute{Swedish Institute of Space Physics, Uppsala, Sweden\\
              \email{andrew.dimmock@irfu.se}
         \and
             Department of Physics, Ben-Gurion University of the Negev, Beer-Sheva, Israel
              \and
              Space and Plasma Physics, Department of Physics and Astronomy, Uppsala University, Uppsala, Sweden
              \and
             Imperial College London, London, UK
              \and
              Swedish Defence Research Agency, Stockholm, Sweden
              \and
             University of Turku, Turku, Finland
              \and
             Departamento de Ciencias Espaciales, Instituto de Geofísica, Universidad Nacional Autónoma de México, Ciudad Universitaria, Ciudad de México, Mexico
             \and
             Mullard Space Science Laboratory, University College London, UK
             \and
             Institute of Experimental and Applied Physics, Kiel University, 24118 Kiel, Germany
             }

   \date{Received May 25, 2023;}

% \abstract{}{}{}{}{} 
% 5 {} token are mandatory
 
  \abstract
   {Solar Orbiter, a mission developed by the European Space Agency, explores in situ plasma across the inner heliosphere while providing remote-sensing observations of the Sun. The mission aims to study the solar wind, but also transient structures such as interplanetary coronal mass ejections and stream interaction regions. These structures often contain a leading shock wave that can differ from other plasma shock waves, such as those around planets. Importantly, the Mach number of these interplanetary shocks is typically low (1-3) compared to planetary bow shocks and most astrophysical shocks. However, our shock survey revealed that on 30 October 2021, Solar Orbiter measured a shock with an Alfv\'en Mach number above 6, which can be considered high in this context.}
   {Our study examines particle observations for the 30 October 2021 shock. The particles provide clear evidence of ion reflection up to several minutes upstream of the shock. Additionally, the magnetic and electric field observations contain complex electromagnetic structures near the shock, and we aim to investigate how they are connected to ion dynamics. The main goal of this study is to advance our understanding of the complex coupling between particles and the shock structure in high Mach number regimes of interplanetary shocks.}
   {We used observations of magnetic and electric fields, probe-spacecraft potential, and thermal and energetic particles to characterize the structure of the shock front and particle dynamics. Furthermore, ion velocity distribution functions were used to study reflected ions and their coupling to the shock. To determine shock parameters and study waves, we used several methods, including cold plasma theory, singular-value decomposition, minimum variance analysis, and shock Rankine-Hugoniot relations. To support the analysis and interpretation of the experimental data, test-particle analysis, and hybrid particle in-cell simulations were used.}
   {The ion velocity distribution functions show clear evidence of particle reflection in the form of backstreaming ions several minutes upstream. The shock structure has complex features at the ramp and whistler precursors. The backstreaming ions may be modulated by the complex shock structure, and the whistler waves are likely driven by gyrating ions in the foot. Supra-thermal ions up to 20 keV were observed, but shock-accelerated particles with energies above this were not.}
   {}


   \maketitle
   
%
%-------------------------------------------------------------------

\section{Introduction}
%% General opening statement
Shock waves are a fundamental plasma phenomenon discovered in numerous diverse plasma environments. Examples of shocks can be encountered close to planets, comets, the solar wind, and across astrophysical environments such as supernova remnants. Shocks are important because they are capable of accelerating particles to remarkably high energies \citep{malkov2011}. For most astrophysical plasma, particle collisions are so rare that their influences can be neglected, thus they are considered collisionless. At collisionless shocks, energy conversion takes place via the interplay between the electromagnetic fields and the particles themselves. A critical shock parameter is the shock obliquity, that is, the angle between the shock normal and the upstream magnetic field, denoted by $\theta_{bn}$. This angle determines whether a shock is quasi-parallel ($\theta_{bn}<45^\circ$) or quasi-perpendicular ($\theta_{bn}>45^\circ$), with important implications for how particles behave across the shock transition. The shock Mach number ($M_A = V_u/V_A$, where $V_A$ is the Alfv\'en speed) is also essential, and although other aspects can be important (e.g., curvature radius or spatial extent and $\beta_i = 2\mu_0 nkT_i/|\mathbf{B}|^2$), $\theta_{bn}$ and $M_A$ determine to a considerable extent how the plasma is processed by the shock front and the formation of its magnetic structure.

%% IP shock motivation
Much of the work performed on shocks within the heliosphere has focused on the physics of the Earth's bow shock because it benefits from an increased number of in situ multispacecraft observations often yielding very high-cadence field and particle measurements. However, there have also been considerable investigations of interplanetary (IP) shocks, that is, shocks in the solar wind that arise due to transients such as interplanetary coronal mass ejections (ICMEs) \citep{hildner1977} and stream interaction regions (SIRs) \citep{richardson2018}. These shocks normally have lower Mach numbers than their planetary counterparts \citep{gosling1983}, larger curvature radii, and are important because they can accelerate particles to high energies ~\citep[e.g.,][and references therein]{perri2022}. However, most of the IP shock studies have either been based on data collected at 1 AU due to the regular presence of spacecraft at the first Lagrange point (e.g., Wind and ACE\footnote{Advanced Composition Explorer (ACE)}), the occasional excursion of magnetospheric missions into the solar wind that happen to encounter an IP shock \citep[e.g.,][]{cohen2019,kajdic2019} as well as events observed by STEREO\footnote{Solar Terrestrial Relations Observatory (STEREO) }\citep{russell2009,blancocano2016} also around 1 AU.

%% ion reflection in general
Multiple physical processes that govern quasi-perpendicular collisionless shocks can be connected to the Mach number, which has led to the derivation of critical Mach numbers \citep{kennel1985}. More appropriate here is the first critical Mach number. When it is exceeded, additional mechanisms other than resistivity (in this case, ion reflection) are required to balance the nonlinear steepening of the shock ramp. These shocks are called supercritical, and their magnetic profile consists of foot, ramp, and overshoot regions. In practice, a fraction of incident ions is reflected by the cross-shock potential toward the upstream direction. Because the ions will gyrate around the upstream field ($\mathbf{B_{us}}$), they will acquire energy from the upstream convective electric field $\mathbf{E_{us} = -V_u \times B_{us}}$ and then will be able to traverse the shock front into the downstream. The gyrating ions are observed directly upstream of the shock ~\citep{Gosling1982GRLEvidence-for-Sp,Paschmann1982GRLObservations-of,Sckopke1983JGEvolution-of-Io}, which means that even at the bow shock, this is typically only for several seconds. However, the time for which a spacecraft can measure this will depend on the shock dynamics, such as whether the shock front is slow or if the shock surface is rippled or nonstationary \citep{johlander2016,madanian2021}.

For more oblique shocks, reflected ions can be observed further upstream, sometimes several minutes before the ramp \citep{Meziane2004, kajdic2017}. From here on, we refer to these separate classes of reflected ions as gyrating (close to the shock) and backstreaming ions (farther away and escaping). In some circumstances, gyrating ions and backstreaming ions can be observed for the same shock \citep{kajdic2017}. Backstreaming ions can also be divided into subsets of ion populations such as field-aligned beams (ions aligned with the magnetic field), diffuse (isotropic characteristics), and intermediate ions that have characteristics of both beams and diffuse ions \citep{paschmann1981}. These classes of backstreaming ions are particularly important to the present study. Furthermore, the physics of ion reflection at collisionless shocks has been examined in the greatest detail at the terrestrial bow shock; as expected, a wealth of literature is available on this topic, often complemented by theoretical and numerical modeling that in the last decades was proven to be an invaluable tool to connect theoretical knowledge and spacecraft observations \citep{Leroy1981, Gedalin1996JGIon-Reflection-,Gedalin2008JGRDistribution-of,Kong2017, Trotta2019, Preisser2020}.

%%%

The signature of ion reflection from spacecraft measurements is populations of ions upstream of the shock ramp that are separate from the solar wind. These are typically clear in the observed ion velocity distribution function (VDF). In the shock frame, distributions that have reflected ions should contain the reflected component with velocities that are directed antiparallel to the shock normal (i.e., moving away from the shock ramp) and the incident flow moving toward the shock. In reality, this feature can appear to be modulated by more complex processes at the shock front. Examples of this could be rapid back-and-forth motion of the shock, rippling \citep{johlander2016}, or shock reformation when the Mach number is sufficiently high \citep{krasnoselskikh2002,dimmock2019}. 

A slow shock speed (i.e., providing a longer transit time across the shock) can allow for these features to be resolved in greater detail or not at all when the shock speeds are fast enough. Other processes can also be connected to reflected ions, such as waves and instabilities \citep{wu1984,lalti2022}. This sampling issue is particularly prominent at IP shocks because the spacecraft traverses the shock front with typical speeds that are an order of magnitude higher than that of the Earth's bow shock. Thus, capturing particle distribution functions with a cadence of several seconds provides a limitation on how well shock features and structures can be resolved in these measurements.

%% Ion reflection at IP shocks, general description
Ion reflection at IP shocks is much less frequently studied and less well understood than at the Earth's bow shock. One benefit of studying ion dynamics at IP shocks is that their field line connection to the upstream occurs for more extended periods compared to the bow shock (one day at 1AU compared to 10 minutes at the bow shock) \citep{gosling1983,Lario2022}. As a result, it is commonly acknowledged that acceleration processes will be more evolved than in bow shocks, where the early stages are observed \citep{gosling1983}. Thus, it is claimed that we would typically envision higher-energy diffuse ion populations. Some studies have reported the existence of multiple ion populations at IP shocks, however \citep[see][]{kajdic2017,zho2020}, demonstrating that this is not always the case. A clear characterization of the features of reflected ions upstream of IP shocks is therefore still not fully settled.

%% Kajdic 2017 
\citet{kajdic2017} studied various types of ion populations at an IP shock encountered on 8 October 2013 using ARTEMIS and ACE observations. The shock had a fairly high Mach number (4.9-5.7) and was oblique ($\theta_{bn} = 47-61^\circ$) when measured at both spacecraft locations. Using Artemis, the authors reported the first observations of multiple suprathermal ion populations at an IP shock. The observed ion populations were a mixture of field-aligned beams, gyrating ions, intermediate ions, and diffuse ions. The energy ranges of the suprathermal ions were notably large; the field-aligned beams were approximately 20 keV, whereas the diffuse ions ranged up to 200 keV (spacecraft frame). The authors suggested that IP shocks remain connected to the upstream magnetic field lines for much longer than planetary bow shocks, permitting shock-drift acceleration and shock surfing \citep{lever2001} to function for longer and can result in more highly energetic ions.

%% Cohen 2019
On rare occasions, ion reflection at IP shocks has been observed with magnetospheric missions with extremely high-cadence plasma measurements. For example, the MMS spacecraft observed an IP shock on 8 January 2018 that was studied in detail by multiple authors \citep{cohen2019,hanson2020}. \citet{cohen2019} showed conclusive evidence of near specularly reflected (gyrating) ions close to the shock front, which is expected because the shock exceeded the first critical Mach number. In their case, the spacecraft was able to resolve these features because it operated in burst mode, which may not be feasible with other spacecraft. \citet{hanson2020} studied the same shock and reported backstreaming ions several minutes upstream of the ramp. The energies of these ions were around 2-7 keV, and the authors concluded with the help of test-particle analysis that these ions were accelerated by the shock-drift acceleration process.

%% Zho20200
Using ARTEMIS, \citet{zho2020} studied energetic ions reflected from IP shocks. In this study, they observed an energy dispersion in which higher energies were observed farther upstream of the shock. In addition, energetic ions around 10 keV were discovered 10 minutes both upstream and downstream of the shock. Similar to \citet{kajdic2017}, the authors proposed that multiple ion populations are present, namely a lower-energy component closer to the ramp (1-4 keV), and a more energetic population (4-25 keV) that can be located farther from the shock ramp. Interestingly, the authors suggested that energetic ions upstream contained escaping ions, even though the geometry exceeded $60^\circ$. These ions were also observed earlier at a high Mach number quasi-perpendicular terrestrial bow shock~\citep{Kucharek2004AGOn-the-Origin-o}.

%% Motivation for the current paper
The recent Solar Orbiter mission provides a considerable advancement in the capability to study IP shocks by delivering continual high-quality measurements of the solar wind across heliospheric distances between around 0.3 AU to 1 AU. This paper uses Solar Orbiter measurements to survey IP shocks with the aim to identify high Mach number events that exhibit signs of ion reflection. To do this, we investigated an IP shock observed on 30 October 2021 that revealed rare characteristics of ion reflection as well as other interesting features. To aid the study, test-particle analysis and kinetic simulations are performed to obtain a more comprehensive understanding of this event.

\section{Data sources and event selection}
\subsection{Solar Orbiter data and instrumentation}
We used observations from the Solar Orbiter \citep{muller2020} spacecraft spanning the period between 15 April 2020 and 31 August 2022. The fluxgate magnetometer instrument (MAG) \citep{horbury2020} was used to characterize the magnetic field structure of the shock front and the properties of the upstream and downstream regions. The probe-spacecraft potential ($ScPot$) was measured by the radio and plasma wave experiment (RPW) \citep{maksimovic2020} and was calibrated to estimate the electron density \citep{khotyaintsev2021}, which we refer to as $N_e$ from this point on. This measurement provides a useful dataset for studying density variations within the shock ramp because its temporal resolution is high, which is required to capture variations on timescales of seconds or shorter. 

The solar wind analyzer (SWA) instrument \citep{owen2020} proton alpha sensor (SWA-PAS) delivers ion VDFs and ion moments every 4 seconds with an energy range up to around 8 keV. Here, ground-based moments were determined from the proton peak in the SWA-PAS VDFs, but it is important to mention that the proton peak can be difficult to distinguish in certain situations, and alphas may sometimes affect the moment calculations. To identify higher-energy ions, the Energetic Particle Detector (EPD) \citep{rodriguez2020} was used.

In this paper, we use the RTN (radial, tangential, normal) coordinate system unless stated otherwise. The CDF file versions used for the detailed analysis were MAG (1), SWA-PAS(3), RPW (1), and EPD-STEP (1).

\subsection{Shock survey and event selection}
The goal of this study is to investigate the signatures of ion reflection at quasi-perpendicular IP shocks. We therefore initially performed a thorough survey of the Solar Orbiter data to identify suitable events. The survey was performed using an automated search from the beginning of the mission until 31 August 2022. The method was adopted from the IP shock database maintained by the University of Helsinki \citep[see][]{kilpua2015}. The automated shock search requires both MAG and SWA-PAS data. We provide the outcome from this survey in Table \ref{tab:shock_list} in Appendix \ref{appendix:shock list}. Although SWA-PAS and MAG data are required to study ion reflection, we also included events here that were identified by eye when these observations were not available. In total, we found 47 IP shocks, 13 of which lacked either MAG or SWA-PAS. 

\begin{figure*}
   \centering
   \includegraphics[width=13cm]{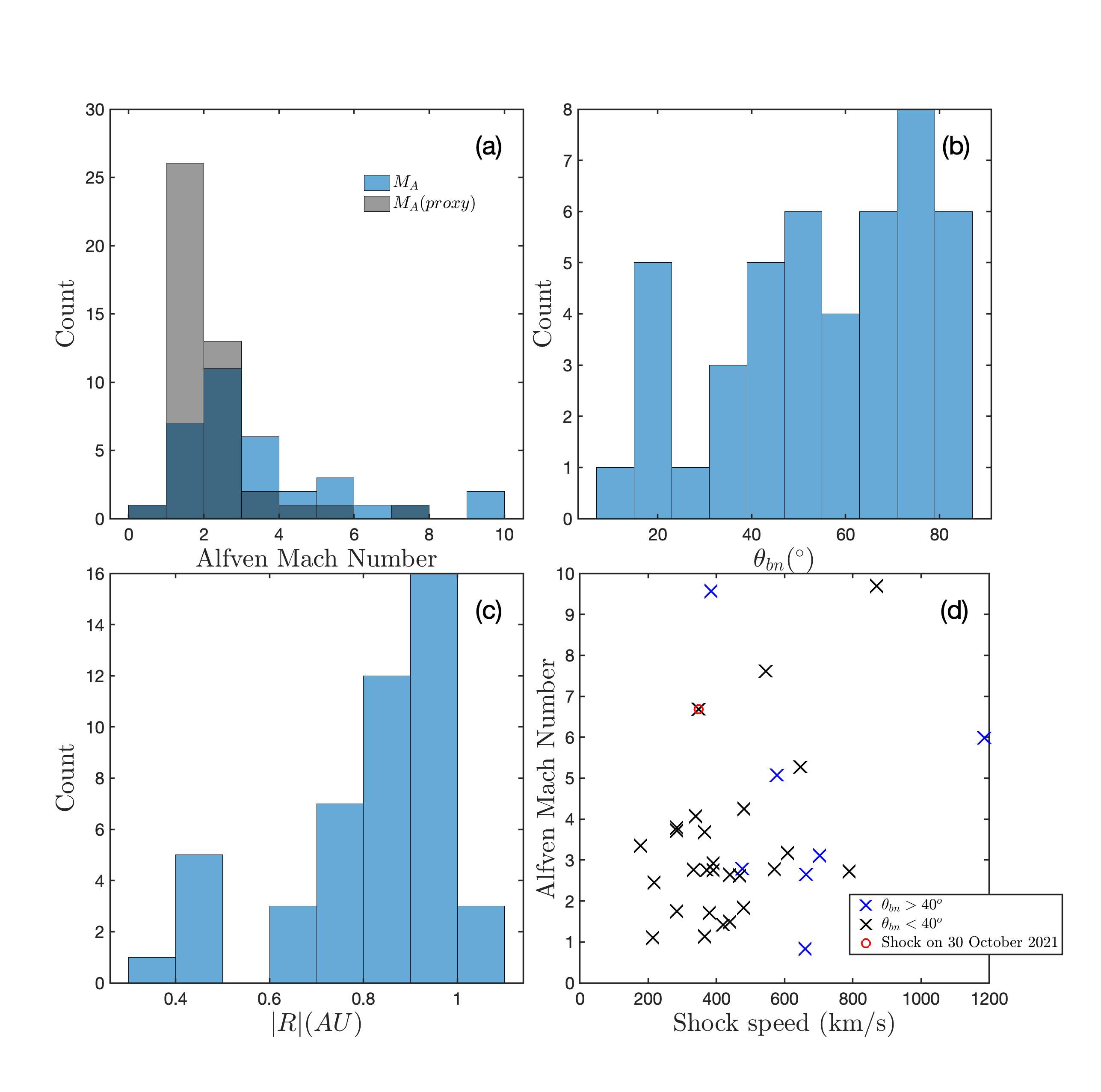}
      \caption{Solar Orbiter IP shock properties. Panel (a) shows $M_A$ both computed from the MAG and SWA-PAS observations and $M_A$ estimated from $B$ according to Equation \ref{proxy}. The shock geometry is shown in panel (b) and was computed from mixed-mode 3 if all data were available or from magnetic coplanarity if only MAG was available. The geocentric distance of each shock is shown in panel (c) and the compression ratios in panel (d). Missing entries signify lack of data.}
         \label{fig:solo_sdb_histo}
\end{figure*}

Two critical shock parameters for the presence of ion reflection are the Mach number and $\theta_{bn}$ \citep{kennel1985}. For this study, we targeted high Mach number shocks ($M_A>5$) that are quasi-perpendicular. Figure \ref{fig:solo_sdb_histo}a-b shows a histogram of these parameters for the Solar Orbiter event list, showing that although most shocks have $\theta_{bn}>40^\circ$, the Alfv\'en Mach number is typically low. Ion reflection is therefore expected to be rare. Panel (d) also includes the heliocentric distance ($|R|$), showing that although some shocks were detected close to perihelion, most were when $|R|>0.6$.

Another key parameter for resolving complex features of the shock related to ion reflection is the speed of the shock relative to the spacecraft. Slower shock speeds allow resolving key features of the shock, especially because ion distributions have a cadence of about 4 seconds, which can be comparable to or greater than the typical temporal scales of these features. Ideally, this should not be too fast, even though the Mach number needs to be high. A histogram of the shock speed (see Appendix \ref{app:shock_cal} for how this is calculated) is shown in Figure \ref{fig:solo_sdb_histo}d. The color of each marker indicates when the shock obliquity is above (black) or below (blue) 40$^\circ$. Although quasi-perpendicular shocks are conventionally defined as having $\theta_{bn}>45^\circ$, we opted for a slightly lower value of $40^\circ$ to account for the uncertainty in calculating $\theta_{bn}$ and to avoid ruling out shocks that are within a few degrees of this.

Figure \ref{fig:solo_sdb_histo}d highlights an event that matches our criteria with a moderate speed, quasi-perpendicular ($\theta_{bn}\sim45$), and $M_A=6.7$. The event exhibited signatures of ion reflection. The Solar Orbiter observations of this shock crossing are the main focus of this paper and are analyzed in detail in the following section. It is important to note that there are seven shock crossings when $M_A>5$, four of which correspond to $\theta_{bn}>40^\circ$. We visually inspected each of these shocks, and only the one mentioned above demonstrated clear signatures of ion reflection in SWA-PAS when the data were of sufficient quality for a detailed analysis. This event is studied in detail and is complemented with theory and simulations below to obtain a more complete understanding of the event.

\section{Case study: IP shock on 30 October 2021}
\subsection{Large-scale structure}
\begin{figure*}
   \centering
   \includegraphics[width=18cm]{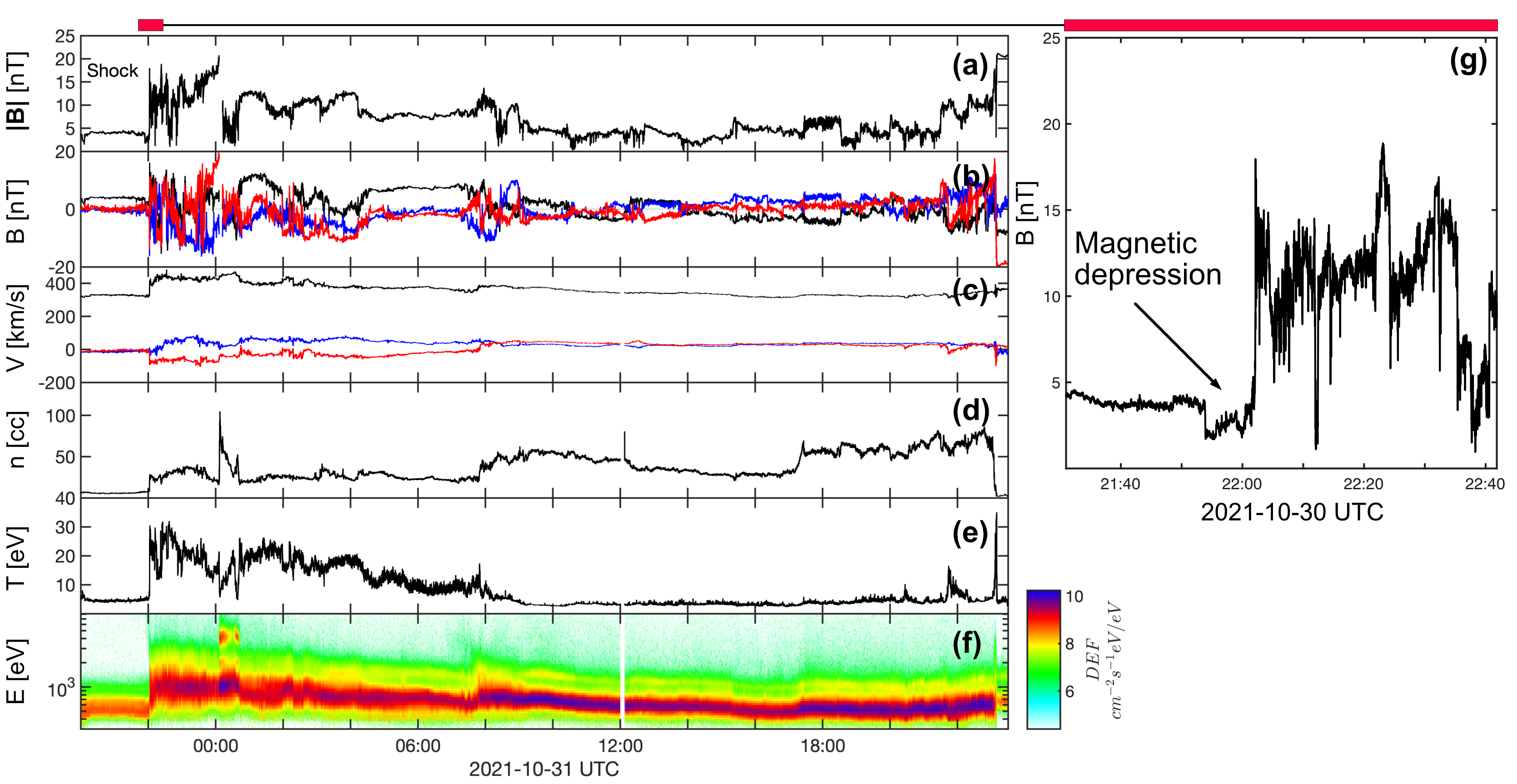}
      \caption{ICME observed by Solar Orbiter between 30-31 October 2021. Panels (a-e) show $|B|$, $B$, $V$, $n$, $T$, respectively and the bottom panel (f) is the DEF.}
         \label{fig:shock_icme}
\end{figure*}

From our survey, we focus on an event between 30 and 31 October 2021 when Solar Orbiter was at a distance of 0.8 AU. As shown in Figure \ref{fig:shock_icme}, Solar Orbiter observed an ICME, where panels a-e show $|\mathbf{B}|$, $\mathbf{B}$, $\mathbf{V}$, $n$, and $T$, respectively. The bottom panel, panel (f), displays the omnidirectional ion differential energy flux (DEF).
The shock occurred at 22:02 UT on 30 October. This was identified by the increase in the magnetic field, ion velocity, ion temperature, ion density, and ion energy. The region directly downstream of the shock contained a high level of magnetic field turbulence until the end of the day, when it decreased significantly on 1 November 2021. We interpret the disturbed region as a sheath region and the subsequent several hours as a magnetic cloud. Therefore, this interval likely corresponds to a slow ICME.

Analysis of the larger-scale structure is important as it sheds light on how the high $M_A$ was reached, which is a key requirement for ion reflection. In panel g, this seems to arise from the locally low upstream magnetic field. Crucially, in panel (g) in Figure \ref{fig:shock_icme}, it appears that a localized depression in the magnetic field before the arrival of the shock also contributes to this low magnetic field and slow Alfv\'en speed. This feature may also imply that this shock may have a locally high $M_A$ and may not reflect the value for the global shock surface. This further highlights the rarity of high $M_A$ IP shocks: The conditions are not easily satisfied. Furthermore, it emphasizes why this event deserves a thorough analysis.

\subsection{Macroscale shock features and parameters}
\begin{figure*}
   \centering
   \includegraphics[width=18cm]{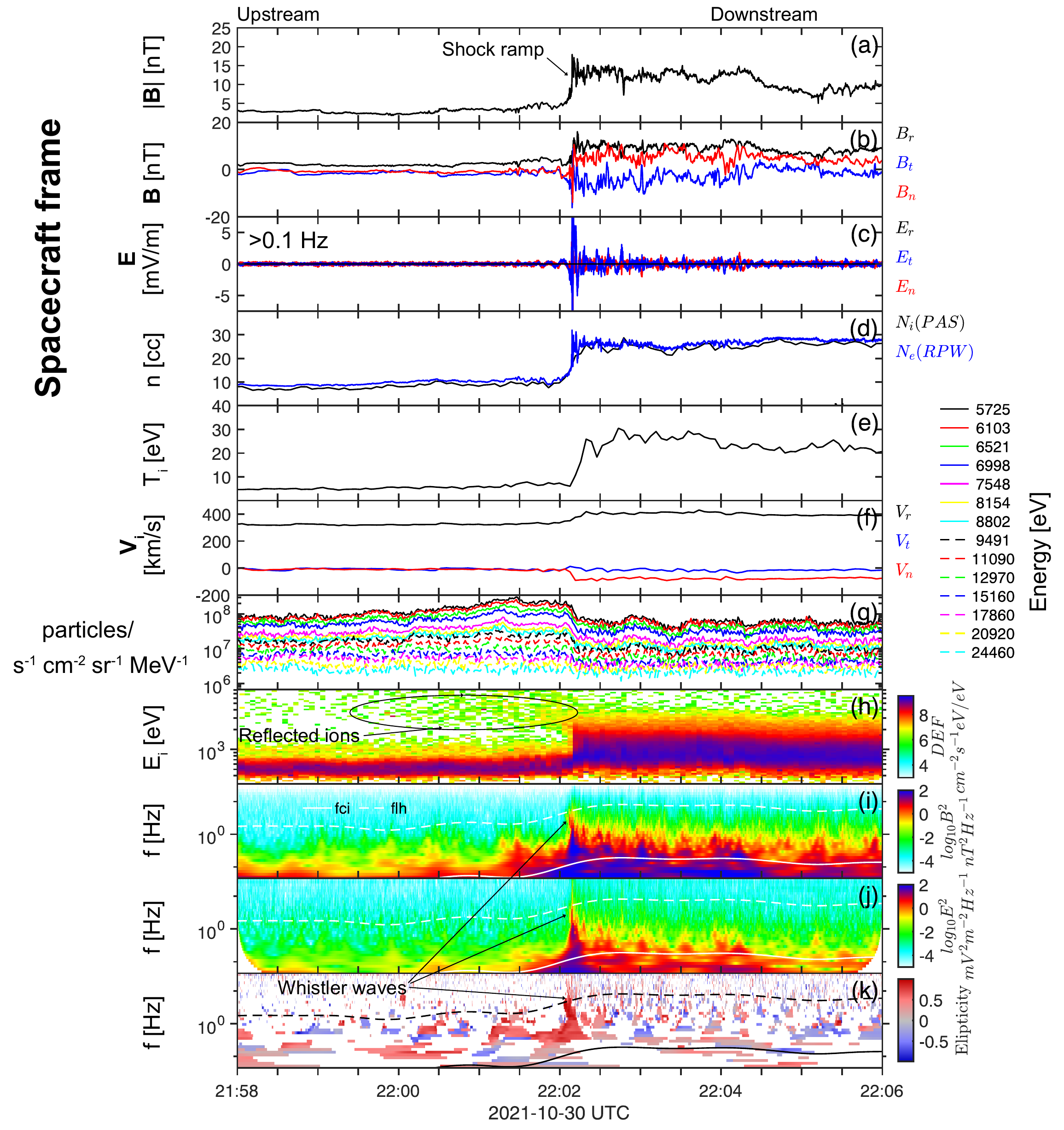}
      \caption{Overview of the IP shock observed by Solar Orbiter on 30 October 2021. Panels a-k from top to bottom represent $\mathbf{|B|}$, $\mathbf{B_{rtn}}$, $\mathbf{E_{rtn}}$, $n_{i,e}$, $T_i$, $\mathbf{V_{rtn}}$, EPD ion energy, SWA-PAS differential energy flux, $\mathbf{B}$ wavelet spectra, $\mathbf{E}$ wavelet spectra, and the ellipticity of $\mathbf{B}$. The shock crossing occurs around 22:02 UTC and is marked by the sudden change in all the plotted quantities.}
         \label{fig:shock_overview}
\end{figure*}

Now that the context for the high $M_A$ has been provided, we proceed with the analysis of the shock front itself. Figure \ref{fig:shock_overview} shows an overview of Solar Orbiter MAG, RPW, SWA-PAS, and EPD-STEP observations during the shock crossing around 22:02 UTC. All quantities are in the spacecraft frame and RTN coordinates. Panels a-c show the magnetic and electric fields, and panels (d) to (f) show the density, temperature, and velocity, respectively. Particle intensities in the magnet channel from EPD-STEP data are plotted in panel g for various bins ranging from 5.7 keV to 24.4 keV. The DEF from SWA-PAS is shown in panel (h) and extends to 8 keV. Wavelet spectrograms for magnetic and electric fields are located in panels (i) to (j), and the ellipticity of $\mathbf{B}$ is shown in the bottom panel (k). There is value to showing quantities in both the spacecraft frame and shock rest frame because additional information is provided in each quantity, and some shock features are clearer. Therefore, we separated the observations in each frame and later provide observations in the shock rest frame.

The magnetic field profile of the shock contains unmistakable foot, ramp, and overshoot characteristics that are consistent with supercritical quasi-perpendicular collisionless shocks. To calculate the shock normal direction, we used the mixed-mode coplanarity method \citep{paschmann1998}, which resulted in $\hat{\mathbf{n}} = [0.64, -0.04, -0.77]$. The upstream magnetic field is $\mathbf{B_u} = [1.99, -1.39, -0.60]$ nT, which provides a shock geometry of $\theta_{bn} = 44^\circ$. This was also compared to other methods, and we found that the shock geometry may change by up to $\sim 30^\circ$, which is likely due to the variability associated with the event ~\citep{Trotta2022b}. The shock speed was estimated to be 348 km/s, which gives $M_A = 6.7$ and $M = 2.5$ (assuming $T_e = 14$ eV). The parameters for this shock are documented in Table \ref{tab:shock_params}.

\begin{table}
\caption{Parameters for the shock crossing.}
\label{tab:shock_params}
\centering
\begin{tabular}{l l}
\hline
Parameter  & Value  \\
\hline
Date  & 30 October 2021  \\
Time of shock ramp (UTC)  & 22:02:09   \\
$\theta_{bn}$$^{a}$ [$^\circ$] & 44\\
Alfv\'en Mach number$^{a}$ $M_A$  & 6.7\\
Fast Mode Mach number$^{a}$ $M_f$  & 2.5\\
$\mathbf{\hat{n}}_{mx3}$ (mixed mode 3) & [0.64, -0.04, -0.77]\\
$\mathbf{\hat{n}}_{mc}$ (magnetic coplanarity) & [0.47, -0.56, -0.68]\\
$\mathbf{\hat{n}}_{mva}$ (minimum variance) & [0.80, -0.20, -0.57]\\
Shock speed$^{a}$ [km/s] & 348\\
Solar wind speed $V_u$ [km/s] & 321\\
Magnetic field $B_u$ [nT] & 2.7\\
Magnetic compression ratio $(B_u/B_d)$ & 3.6\\
Ion Temperature $T_{iu}$ [eV]  & 5.0\\
Density (ion, electron) $n_u$ [cm$^{-3}$]  & 7.3, 8.9\\
Density compression ratio $(n_u/n_d)$ & 3.5\\
Ion plasma $\beta_{iu}$  & 2.3\\
Upstream window $\Delta_u$ & 21:58:00 - 22:00:00\\
Downstream window $\Delta_d$ & 22:05:09 - 22:08:06\\
\hline
\multicolumn{2}{l}{$^{a}$Based on mixed mode 3 shock normal $\mathbf{\hat{n}}_{mx3}$.} \\
\multicolumn{2}{l}{$^{b}$Assuming $T_e=14$ eV.}
\end{tabular}
\end{table}

It is plausible that the shock is nearer to quasi-perpendicular due to the absence of a clear foreshock and accompanying ultra low-frequency (ULF) waves that are typically observed upstream of quasi-parallel shocks \citep{eastwood2005}. Although quasi-perpendicular shocks have been observed with waves \citep{blancocano2016} with frequencies comparable to those in the foreshock, the magnetic structure of the shock and the associated features of the omnidirectional energy spectrogram for this event are more consistent with a quasi-perpendicular geometry. This obliquity is important because shocks with these properties may exhibit multiple types of reflected ion populations.

The SWA-PAS moments plotted in panels (d) to (f) of Figure \ref{fig:shock_overview} show the density, temperature, and velocity changes across the shock front. The electron density from RPW is also included in panel (d) and agrees excellently with SWA-PAS. This measurement is useful because it enables resolving higher-density structures. The DEF in panel (h) shows a population of ions upstream of the shock with energies significantly above the solar wind, which is labeled in the figure as reflected ions. Measurements by EPD indicate that this population extends into the suprathermal range up to around 15 keV. This feature might suggest that the reflected ions reach suprathermal energies, which could be expected for an oblique supercritical shock. Interestingly, the EPD data show a decrease in particle intensity behind the shock ramp, meaning that the suprathermal ions upstream do not reach the downstream. On the other hand, the limited field of view of the EPD instrument may have an impact here, which is a caveat \citep{Wimmer-Schweingruber2021}.

The magnetic field spectrogram in panel (i) and the ellipticity in panel (k) do not show any wave activity far upstream. Thus, we conclude here that there are no indications of any developed foreshock, which would be expected from a more perpendicular shock geometry. On the other hand, there is a small wave packet near the shock ramp that is localized at the shock foot. It is right-handed and circularly polarized and may signify the presence of whistler precursors that are commonly observed upstream of collisionless shocks \citep{wu1984,balikhin1997,wilson2012,dimmock2013,lalti2022}. These waves are relevant to this study as they can be generated by reflected ions, and they therefore also need to be analyzed.

%% Shock observations in shock frame
\subsection{Particle dynamics in the shock frame}
In this section, the measurements are analyzed in the shock frame and shock coordinates. This can provide further information on the behavior of the particles upstream of the shock and also further evaluate the quality of the shock normal. 

\begin{figure*}
   \centering
   \includegraphics[width=12cm]{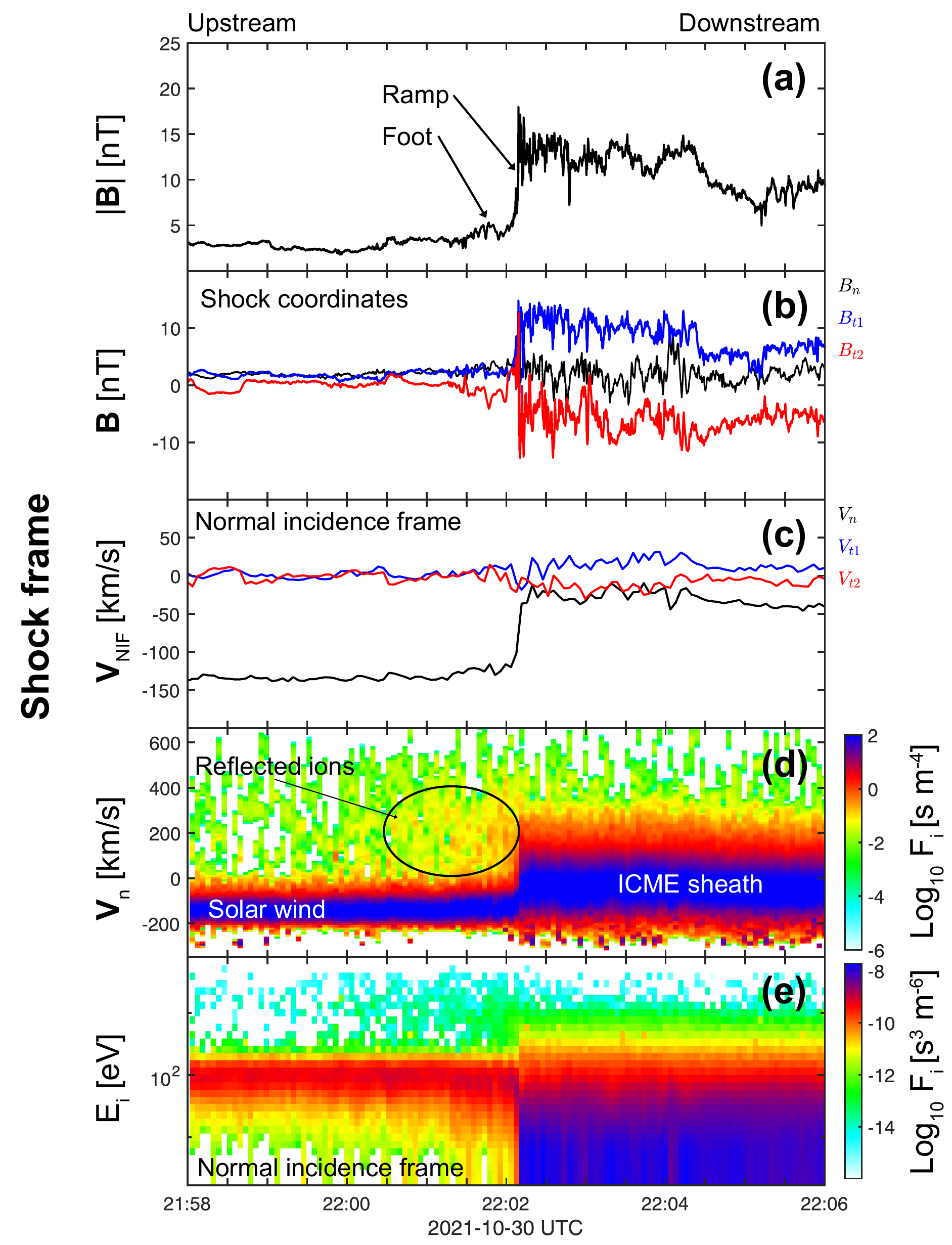}
      \caption{Solar Orbiter observations of the shock crossing transformed into shock coordinates ($\mathbf{\hat n}$, $\mathbf{\hat t_1}$, $\mathbf{\hat t_2}$) and the NI frame.}
         \label{fig:shock_frame}
   \end{figure*}

We plot in Figure \ref{fig:shock_frame} the shock crossing, but in shock coordinates and the normal incidence (NI) frame, which is the shock frame in which the upstream plasma flow is along the shock normal. The shock frame is calculated as $\mathbf{\hat n}$, $\mathbf{\hat t_2}$ , and $\mathbf{\hat t_1}$, where $\mathbf{\hat t_2} = \mathbf{\hat n} \times \mathbf{B_u} / |\mathbf{\hat n} \times \mathbf{B_u}|$ and $\mathbf{\hat t_1} = \mathbf{\hat t_2} \times \mathbf{\hat n,}$ and the NI frame velocity is calculated according to $\mathbf{V_{NIF}} = \mathbf{V_{us}} - (\mathbf{V_{us}} \cdot \mathbf{\hat n} - V_{sh})\mathbf{\hat n}$.
   
Panels (a) to (b) are the magnetic field in shock coordinates showing a minimum offset of $B_n$ (black trace in panel b) that is indicative of an accurate shock normal direction. The slight oscillation during the ramp is addressed below. Panel (c) shows the ion velocity in the NI frame. It is relatively low and further demonstrates that the low upstream magnetic field is required to obtain such a large $M_A$. We plot in panel (d) the ion phase space density reduced along the shock normal direction, and panel (e) shows the omnidirectional phase-space density in the NI frame. Panel (d) reveals a population of ions before the shock ramp that extends for minutes into the upstream direction. As a result of this, ions are present at both negative and positive shock normal velocities, which is unambiguous evidence of ion reflection and provides evidence in addition to the data plotted in the spacecraft frame. This signature is also apparent in panel (e) as the population of ions with higher energy than the solar wind beam, which is similar to the population labeled in panel (h) in Figure \ref{fig:shock_overview}.

\begin{figure*}
 \centering
            \includegraphics[width=18cm]{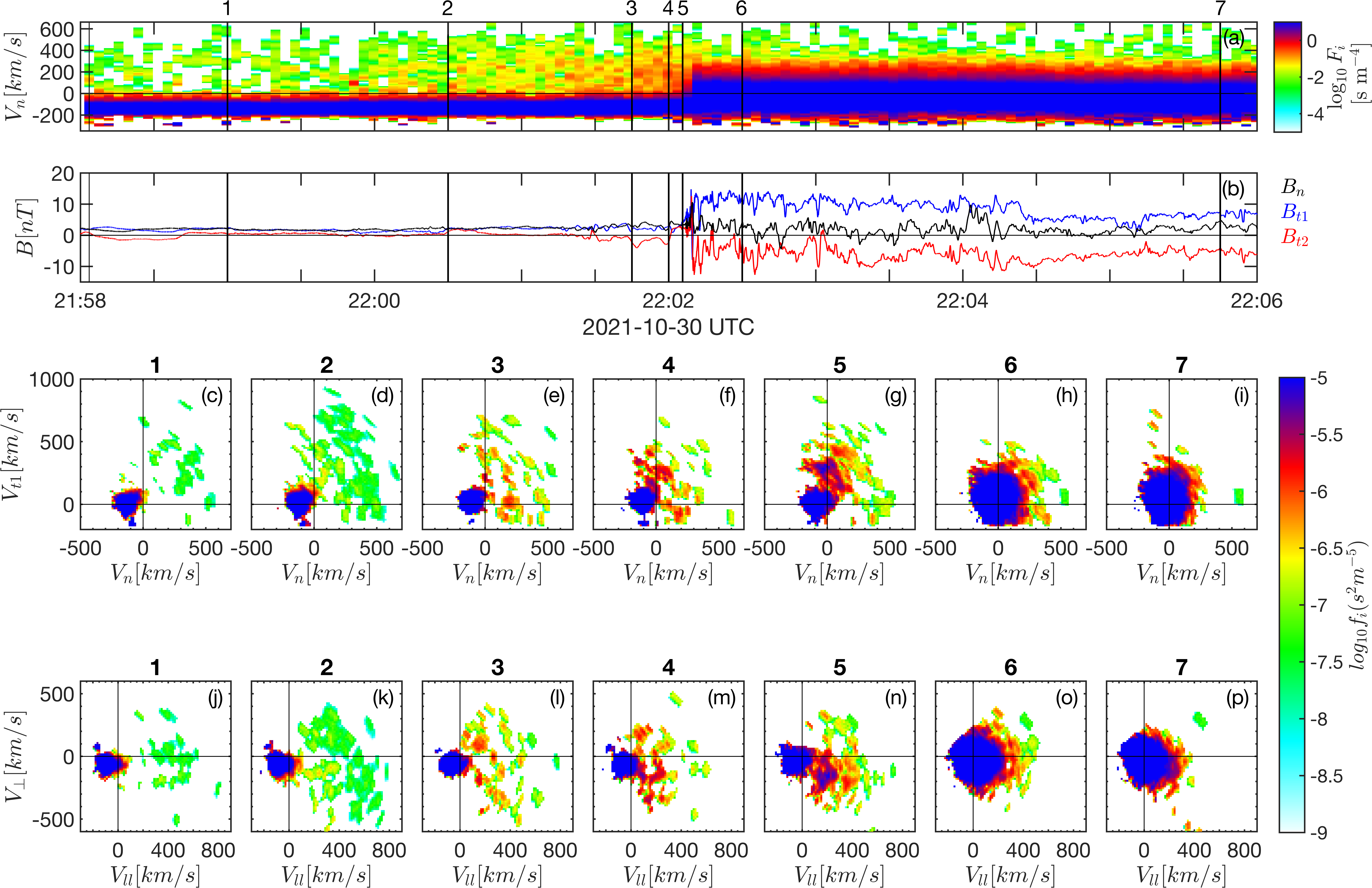}
      \caption{Evolution of ion VDFs across the shock. Panel (a) shows a reduced distribution of the phase-space density along the shock normal, and we plot in panel (b) the magnetic field in shock coordinates. The bottom panels show ion VDFs in the $\mathbf{\hat n}$ - $\mathbf{t_1}$ plane (panels c to i) and $\mathbf{\hat B_{\parallel}}$ - $\mathbf{\hat B_{\perp}}$ at various locations from upstream to downstream of the shock front. $\mathbf{\hat B_{\parallel}}$ refers to $\mathbf{B_u}$. An ion population (in all the VDFs) that is separate from the solar wind beam is noticeable.}
         \label{fig:shock_vdfs}
   \end{figure*}

%\subsubsection{Analysis of ion velocity distribution functions}
The features of the reflected ion can be better characterized by analyzing the 2D ion VDFs around the shock front, where we showed the separate ion populations in the previous figures. In Figure \ref{fig:shock_vdfs}, the ion VDFs are plotted for seven distinct time instances ranging from upstream (1-5) to downstream (6-7), as demonstrated in panels (a) and (b) and as labeled in panels (c) to (i). Each VDF in the top row (panels c to i) is presented in the $\mathbf{\hat n}$ - $\mathbf{t_1}$ plane, and the bottom row is field-aligned.
VDFs 1 and 2 were averaged over five distributions ($\sim 20$ s), and the remaining VDFs are individual distributions. In VDFs 1 to 5, the solar wind population is visible by the narrow core, which is then heated downstream of the shock in VDFs 6 and 7, as revealed by the expanded distribution. VDF 1 shows signatures of ions that are separated from the core and resemble a crescent shape with positive velocities around the core. These are the reflected ions that were visible in the previous figures. In phase space, however, additional features can be extracted.

Closer to the shock, the ion reflection signatures become stronger, and the phase-space density increases. Reflected ions extend several minutes into the upstream direction, which is comfortably outside the magnetic footpoint region, which therefore suggests that these are backstreaming ions. Furthermore, in the lower row of VDFs (panels j to p), the reflected ions are field-aligned, consistent with a backstreaming ion beam. The ion VDFs show reflected ions that are fairly widely dispersed in velocity space and not as focused compared to a well-defined field-aligned beam. Directly upstream of the shock in VDFs 4 and 5, the features become more complex, and there is evidence of additional populations. This may be caused by the more complex magnetic field, by additional ion beams, or by gyrating ions in the foot. We investigated next how these complex features may be related to the shock structure and to the waves in the foot.

\subsection{Magnetic structures in the foot and ramp regions}
It is well established that nonplanar structures such as shock ripples play a significant role in ion reflection \citep{johlander2016,Gedalin1996JGIon-Reflection-}. In addition, reflected ions can generate whistler waves via associated instabilities \citep{wu1984,dimmock2013,lalti2022b}. Therefore, these shock features must be analyzed as they provide invaluable context to the particle observations. Here, we show evidence of both these shock features, and in the discussion below, we explain their significance for the reflected ions that are clear from the ion VDFs.

\subsubsection{Evidence for shock front nonplanarity}
\begin{figure}
   \centering
   \includegraphics[width=\hsize]{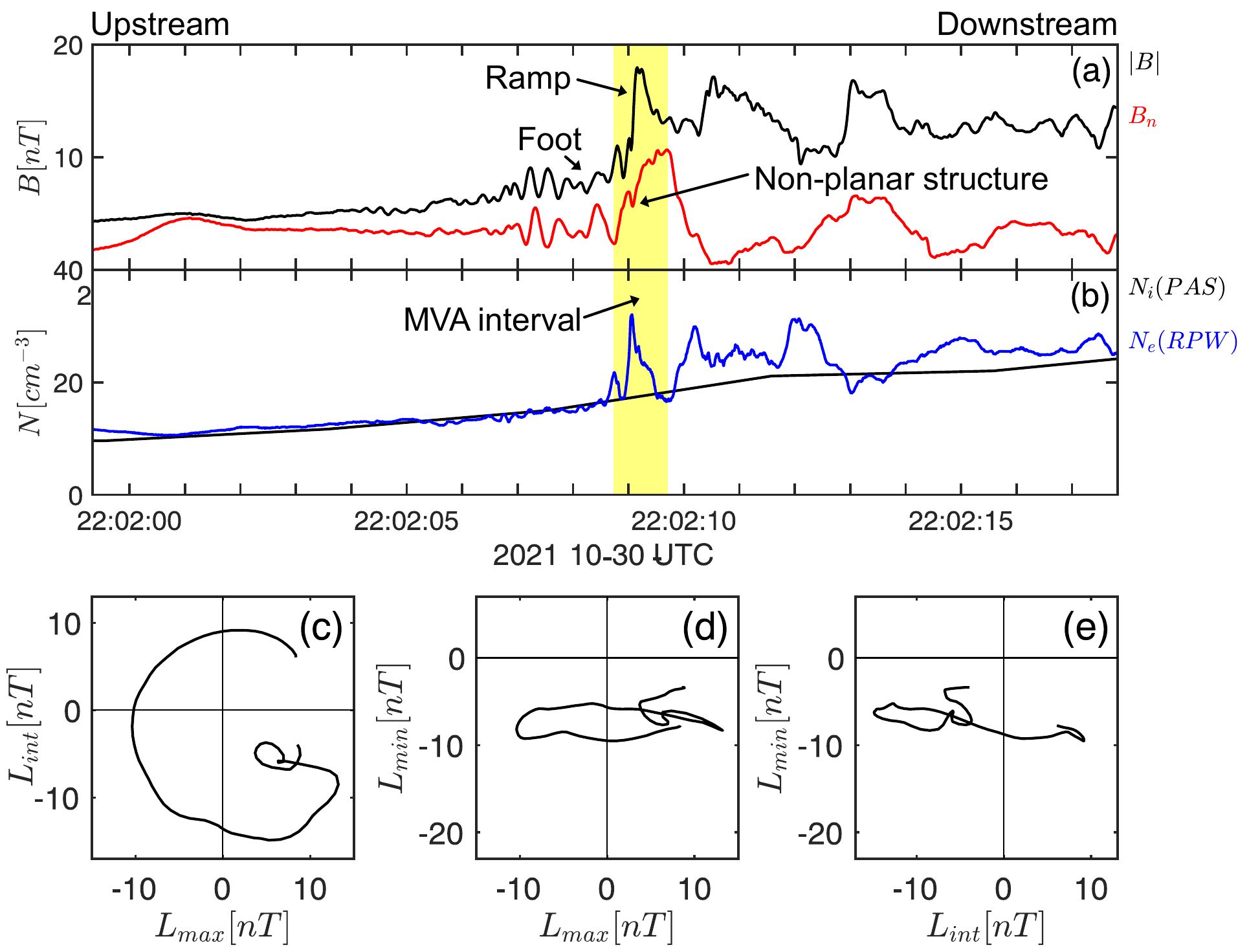}
      \caption{Nonplanar structure inside the shock. Panels (a) and (b) show a zoomed interval of the magnetic field and particle densities across the shock. The bottom panels correspond to hodograms of the structure that is highlighted in yellow in the top panels.}
         \label{fig:shock_nonplanar}
\end{figure}

%% Non-planar structure of the shock front
Observations of nonplanarity require an analysis of a shorter time interval around the shock front. This is presented in Figure \ref{fig:shock_nonplanar}. We plot the magnetic profile of the shock in panel (a), the electron density in panel (b), and hodograms in panels (c) to (e). The hodograms were computed from the time interval within the region highlighted in yellow and demonstrate polarization. They provide indications of the type of shock irregularity.
This timescale shows strong variations in $B_n$ across the shock ramp. For a planar shock, variations in $B_n$ are not expected. This is a strong indication of a substructure that disagrees with the computed shock normal. Using minimum variance analysis (yellow region), we estimated that the angle of the structure to the shock normal is about 25$^\circ$ and 33$^\circ$ to the upstream magnetic field.

To eliminate the possibility that this is caused by a poorly chosen MVA interval, different windows were selected and MVA was performed for each of them. For each MVA, no solution was discovered that did not suggest nonplanar features. The high resolution of the electron density from the RPW instrument also confirms that density variations accompany this nonplanar structure. The SWA-PAS instrument was unable to capture this. The hodograms over this interval reveal that the polarization is either circular or slightly elliptical. These irregularities help to interpret the features in the VDFs shown in Figure \ref{fig:shock_vdfs} and might explain the spread in velocity space of the backstreaming ions.

%% Whistler waves in the shock front
\subsubsection{Whistler precursors in the shock foot}
\begin{figure}
   \centering
   \includegraphics[width=\hsize]{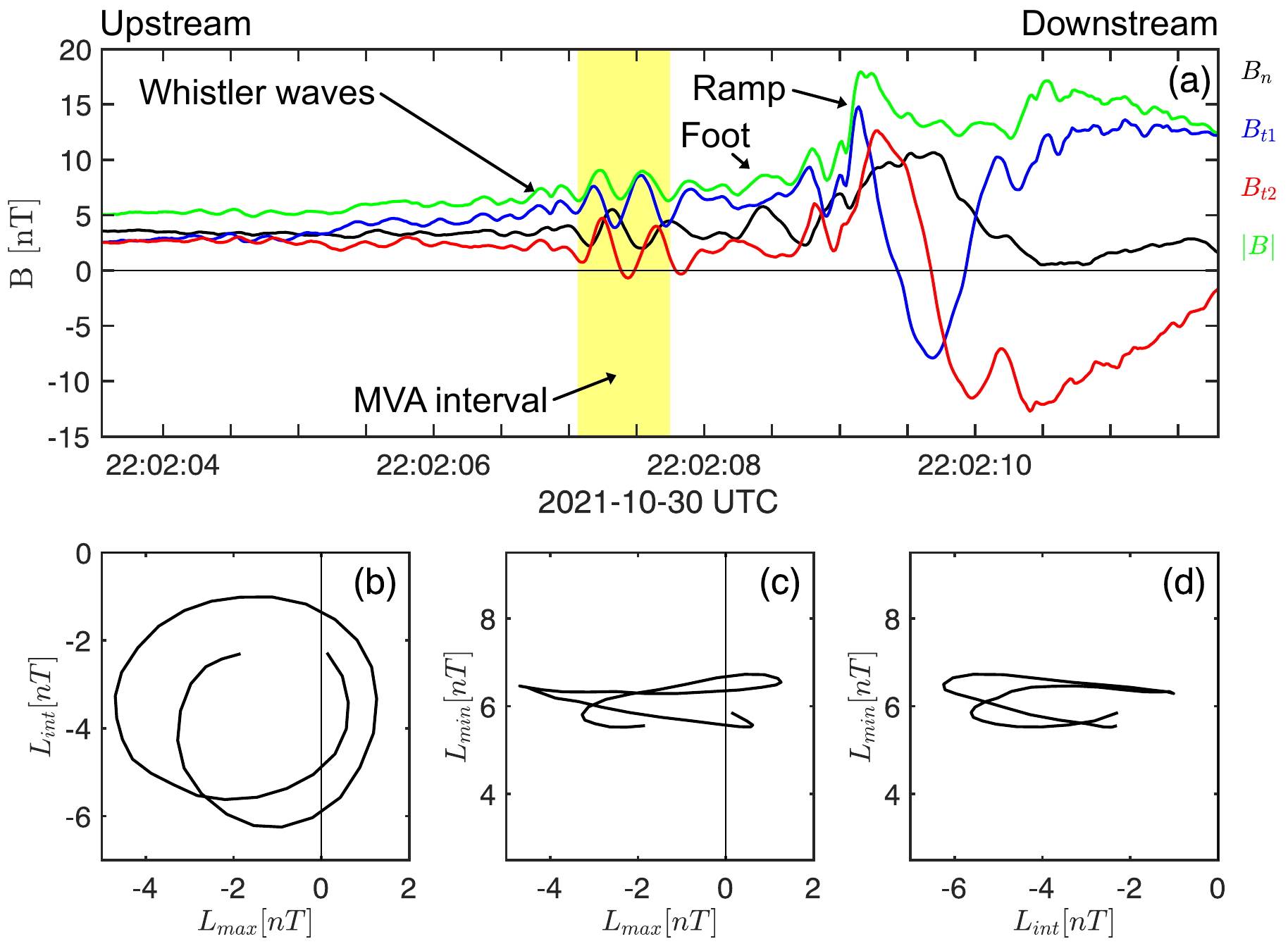}
      \caption{Wave packets in the shock foot region. The top panel shows the magnetic field across the shock front. The bottom panels, panels (b) to (d), are hodograms calculated over the yellow highlighted region in panel (a).}
         \label{fig:shock_foot}
   \end{figure}

To analyze the foot region, we plot another shorter time interval in Figure \ref{fig:shock_foot}, but we now focus on the structures immediately upstream of the ramp. This region (22:02:06-22:02:08) includes notable fluctuations in the magnetic field that may imply the presence of waves (panel a). Similar to before, hodograms of these structures are included to confirm the polarization.
The hodograms from MVA over the region highlighted in yellow suggest that these waves are circularly polarized, and the MVA reveals that their direction to the background magnetic field and shock normal directions are $\theta_{kb} \sim 34^\circ$ and $\theta_{kn} \sim 36^\circ$, respectively. Moreover, the amplitude of the wave packets tends to decay away from the shock front in the upstream direction. These observations are consistent with magnetosonic whistler waves, which are frequently observed upstream of collisionless shocks.

Next, we calculate additional properties of the waves. Combined with those already computed, we identify the most feasible generation mechanism. This is critical because it may indicate a connection to the reflected ions and provide evidence of particle dynamics that could not be resolved. For this situation, we need to rely on single-spacecraft methods and take advantage of the known wave mode. 

A key parameter to determine is the wave number ($k$) because it is used to retrieve the full wave-vector, from which the plasma rest-frame frequency and corresponding wavelength can be obtained. Here, $k$ can be estimated from the dispersion relation of the cold plasma in the whistler wave, as described by \citet{wilson2013} and \citet{wilson2017}. In practice, the following cubic equation was solved for $k$:
\begin{eqnarray}
    0 = \tilde{V}\bar{k}^3 + (cos\theta_{kB} - \tilde{V}\omega_{sc})\bar{k}^2 + \tilde{V}\bar{k} - \tilde{\omega}_{sc},
    \label{eq:disp}
\end{eqnarray}
where $\bar{k} = kc/\omega_{pe}$, $\tilde{\omega}_{sc} = \omega_{sc}/\Omega_{cd}$, $\tilde{V} = V\cos{\theta_{kv}}/V_{Ae}$ and $V_{Ae} = B/\sqrt{\mu_0n_em_e}$. The wave vector direction $\hat{\mathbf{k}}$ can be estimated from minimum variance analysis, and therefore, the angles $\theta_{kb}$ and $\theta_{kv}$ can be computed directly, giving $\theta_{kb} = 34^\circ$ and $\theta_{kv} = 30^\circ$. Solving Equation \ref{eq:disp} gives $k = 7.9552 \times 10^{-5}$ 1/m, which can then be used to calculate the frequency in the plasma rest frame ($\omega_{prf}$) by applying the Doppler shift,
\begin{equation}
    \omega_{sc} = \omega_{prf} + \mathbf{K} \cdot \mathbf{V},
\end{equation}
where $\mathbf{K}$ is the full wave vector. This provides a plasma rest-frame frequency $f_{prf} \sim 6$ Hz and $f_{prf}/f_{lh} \sim 1.3$ compared to $f_{sc} = 2.5$ Hz. 

The parameters of the whistler wave analysis are summarized in Table \ref{tab:waves}. We note that field and plasma properties were collected at the location of the waves and not upstream, as listed in Table \ref{tab:shock_params}. Some discrepancies are therefore expected for these values because the data were collected within the shock foot.
\begin{table}
\caption{Properties of the whistler wave packet used to solve the cold plasma dispersion relation and the resulting parameters.}
\label{tab:waves}
\centering
\begin{tabular}{l l}
\hline
Parameter  & Value  \\
\hline
Time  & 22:02:07   \\
$f_{sc}$ & 2.5 Hz \\
$B$ & 7.7 nT\\
$V$ = [350, 12 -19] & km/s\\
$n_i$ & 15 cm$^{-3}$\\
$\theta_{kb}$ & 34$^\circ$ \\
$\theta_{kv}$ & 30$^\circ$ \\
$\theta_{kn}$ & 36$^\circ$ \\
$|\mathbf{k}|$ & $7.9552 \times 10^{-5}$ 1/m \\
$\lambda$ & 12.6 km\\
$f_{prf}$ & 6 Hz\\
$f_{prf}/f_{lh}$ & 1.3\\
Amplitude (max) & 5.8 nT\\ 
\hline
\end{tabular}
\end{table}
The generation mechanism of the waves is inherently connected to the properties of the waves themselves \citep[see, e.g.,][]{dimmock2013,lalti2022b}. Therefore, the information contained in Table \ref{tab:waves} can be compared to the parameters in other studies to ascertain the most likely generation mechanism. From this, the most likely mechanism is the kinetic cross-stream instability \citep{wu1984,lalti2022}, which is directly connected to ion reflection.

\section{Theory and modeling results}
The Solar Orbiter observations provide a highly localized measurement of the shock front, and although much information can be extracted from these measurements, it is difficult to obtain a global view. In this section, we introduce a theoretical and simulated analysis of comparable shock fronts to address this. First, a test-particle analysis is performed, which enables us to investigate the impact of shock irregularities on the reflected ion beam. Second, because the test-particle method is not self-consistent, we also introduce a hybrid PIC simulation to determine how ripples along the shock front may affect the ion VDFs, and for which situation the VDFs arise that are measured by Solar Orbiter.

\subsection{Test-particle analysis}
Test-particle analyses have been used successfully in the past \citep{Decker1988,Gedalin1996JGIon-Reflection-} to advance our understanding of particle dynamics at collisionless shocks. Particularly, when paired with in situ observations, they can add context to how the trajectory of particles can be altered by nonplanar shock structures. Thus, to complement the Solar Orbiter observations, a test-particle analysis was also performed. We first selected a planar stationary model of the shock, which was set up as the x direction along the shock normal (toward downstream) and y in the noncoplanarity direction. The setup for the planar shock is described in Appendix \ref{appendix: test_particle}. For the analysis in this study, the following values were selected: $M=5$, $\theta_{bn} = 35^\circ$, $D=1/(2M)$, $A_1=0.2$, $A_2 = 2.0$, and $S_{NIF} = 0.3$. In addition, throughout this paper, normalized units are used according to $\mathbf{B} = \mathbf{B} / B_u$, $\mathbf{E} = c\mathbf{E} / V_uB_u$, $\Omega_u = eB_u/m_pc$, and $x = \Omega_ux/V_u$. 

\begin{figure*}
   \centering
\includegraphics[width=14cm]{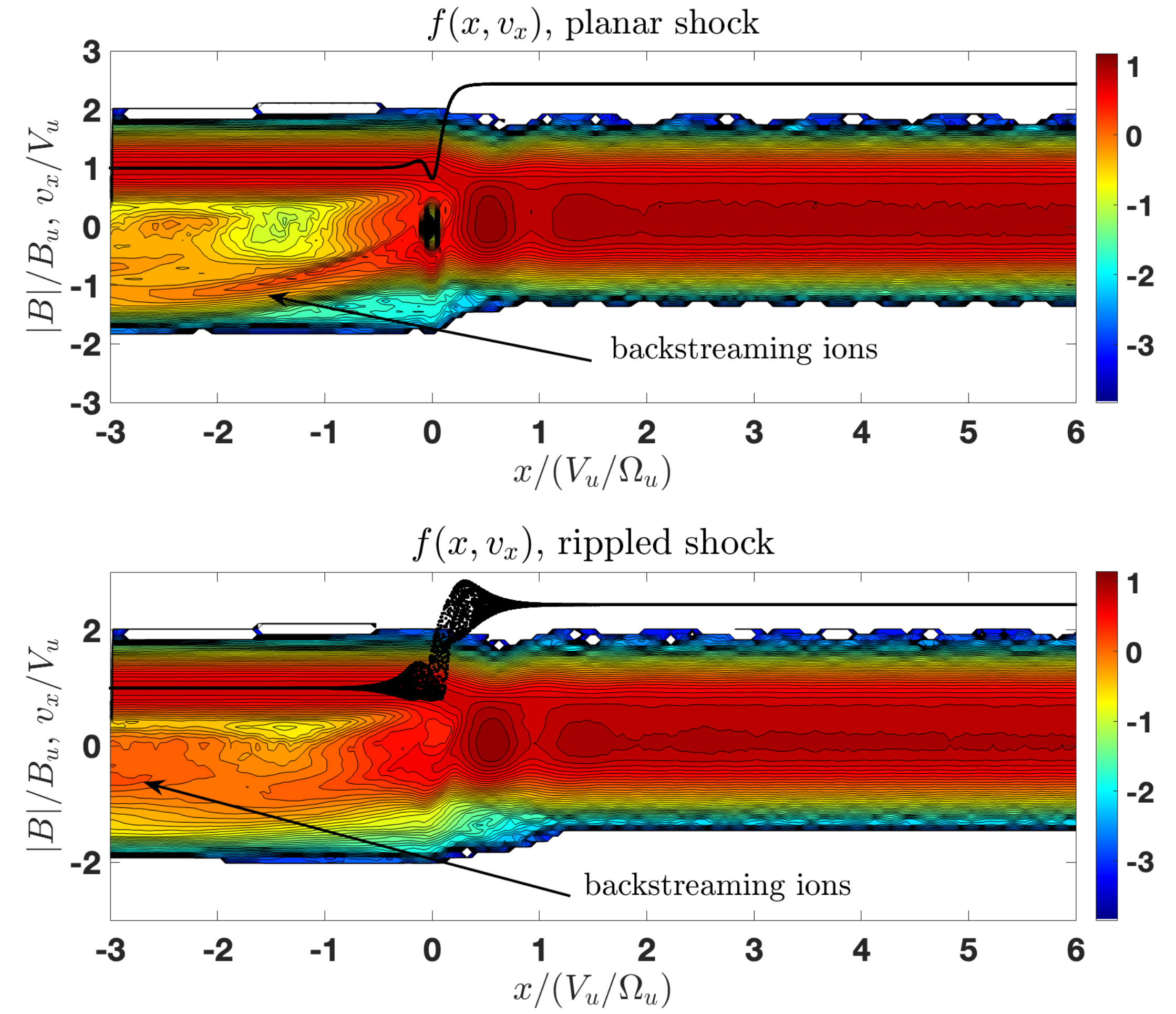}
      \caption{Test-particle analysis of a planar and rippled shock. Top: Reduced distribution $f(x,v_x)$ of ions obtained with a test-particle analysis in a planar shock. The solid black line shows the magnetic field profile. The arrow points to the backstreaming ions. Bottom: Reduced distribution $f(x,v_x)$ of ions obtained with a test-particle analysis in a rippled shock.  The magnetic profile is represented by a ribbon because each ion measures its magnetic field magnitude along its path.}
         \label{fig:test_particle}
   \end{figure*}

The test-particle analysis consisted of 40000 protons that were traced across the shock, and the reduced distribution $f(x,v_x) = \int F(x,v_x,v_y,v_z)dv_ydv_z$ was derived using the staying-time method: The phase space \(x-v_x\) (see Figure~\ref{fig:test_particle}) was covered with a grid. In each cell, we caught each ion as many times as it was found in the cell while moving along its trajectory. The weight \(|v_{x0}|\), where \(v_{x0}\) is the \(x\)-component of the initial ion velocity~\citep{Gedalin2016JGRTransmittedReflectedQuasireflected}, ensured the particle flux conservation. The reduced distribution function was calculated by summation of these weights, multiplied by the number of appearances in the cell, for all traced particles. All particles started at the same position ahead of the ramp. 

Because the observed shock is not exactly planar, it is of interest to compare the test-particle analysis in the planar shock with additional test-particle analyses for a similar but rippled shock. For the setup of the rippled shock~\citep{Gedalin2023ApJScatteringIonsRippleda}, we refer to Appendix \ref{appendix: test_particle}. The tracing was performed by retaining the same initial \(x_{in}\) and \(y_{in}\) coordinates, but randomly choosing the coordinate \(0\leq z_{in}<1\) and the starting time \(0\leq t_{in}<1\), to take both the spatial inhomogeneity along \(z\) and the time dependence into account. 

The reduced distribution function was now averaged over \(z\) and over time. The reduced distributions for both cases are shown in Figure~\ref{fig:test_particle}. 
In the case of the rippled shock, the backstreaming ion beam becomes more widespread, which is prominent. It has been shown that quasi-perpendicular shocks with added rippling produce backstreaming ions, while without rippling, these ions are absent~\citep{Gedalin2023ApJScatteringIonsRippleda}. In this case, the shock is oblique and produces backstreaming ions even without rippling. Rippling increases the number of these ions from 21\% to 26\%, which is not significant. The increased spread in the reflected ion beam for the rippled shock appears consistent with what is observed by Solar Orbiter. The test-particle analysis therefore sheds some light on these observations.

\subsection{Hybrid kinetic particle-in-cell simulations}
\label{sec:hpic_sim}
We also employed kinetic simulations to model the small-scale details of the shock transition, using the hybrid kinetic particle-in-cell (PIC) HYPSI code~\citep[e.g.,][]{Trotta2020a}. In the simulations, protons are modeled as macroparticles and are advanced using the standard PIC method. The electrons, on the other hand, are modeled as a massless, charge-neutralizing fluid with an adiabatic equation of state. The HYPSI code is based on the CAM-CL\footnote{Current advance method and cyclic leapfrog (CAM-CL)} algorithm \citep[][]{Matthews1994}.

The shock was initiated by the injection method, in which the plasma flows in the $x$-direction with a defined (super-Alfv\'enic) velocity $V_\mathrm{in}$. The right-hand boundary of the simulation domain acted as a reflecting wall, and at the left-hand boundary, plasma was injected continuously. The simulation was periodic in the $y$ direction. A shock was created as a consequence of reflection at the wall, and it propagated in the negative $x$ direction. In the simulation frame, the upstream flow is along the shock normal.

In the hybrid simulations, distances were normalized to the ion inertial length $d_i \equiv c/\omega_{pi}$, times to the inverse cyclotron frequency ${\Omega_{ci}}^{-1}$, velocities to the Alfv\'en speed $v_A$ (all referred to the unperturbed upstream state), and the magnetic fields and densities to their unperturbed upstream values, $B_0$ and $n_0$, respectively. The angle between the shock normal and the upstream magnetic field, $\theta_{Bn}$, was 45$^\circ$, with the upstream magnetic field in the $x$-$y$ plane. For the upstream flow velocity, a value of $V_\mathrm{in} = 4.5 v_A$ was chosen, and the resulting Alfv\'enic Mach number of the shock was approximately $M_A \sim 6$, which is compatible with the event observed by Solar Orbiter.

The upstream ion distribution function was an isotropic Maxwellian, and the ion $\beta_i$ was 1. The simulation $x-y$ domain was 512 $\times$ 512 $d_i$. The spatial resolution used was $\Delta x$ = $\Delta y$ = 0.5 $d_i$. The final time for the simulation was 150 $\Omega_{ci}^{-1}$, and the time step for particle {(ion)} advance was $\Delta t_{}$ = 0.01 $\Omega_{ci}^{-1}$. Substepping was used for the magnetic field advance, with an effective time step of $\Delta t_{B} = \Delta t_{}/10$. A small, nonzero resistivity was introduced in the magnetic induction equation. The value of the resistivity was chosen so that there were no excessive fluctuations at the grid scale. The number of particles per cell used was always greater than 300 (upstream) to keep the statistical noise characteristic of PIC simulations to a reasonable level.

\begin{figure*}
   \centering
   \includegraphics[width=16cm]{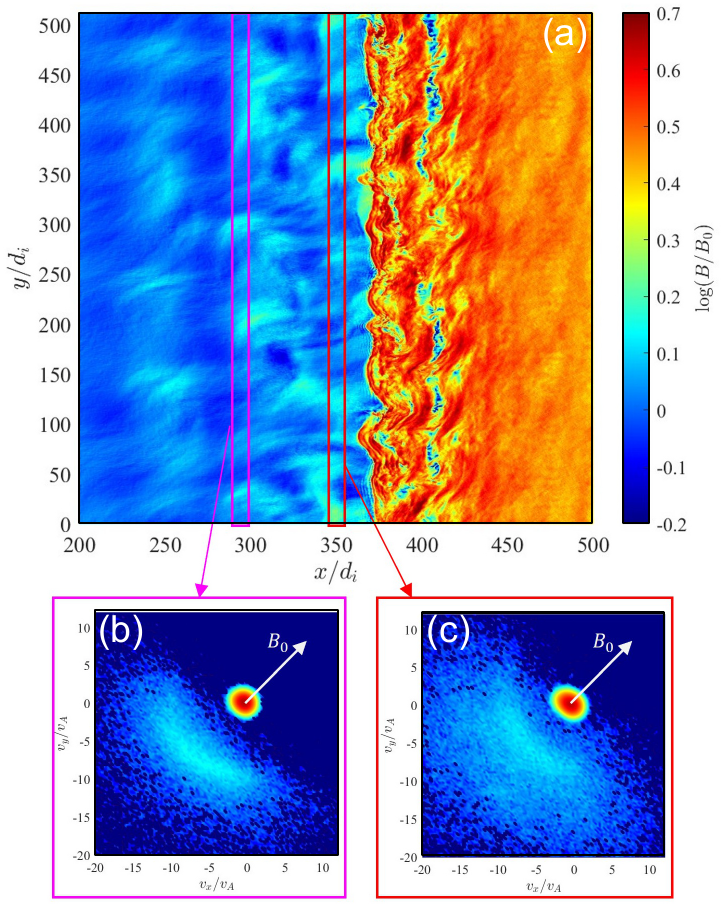}
      \caption{Hybrid PIC simulation for a shock with $\theta_{bn}=45^\circ$ and $M_A \sim 6$. The top panel shows a normalized magnetic field ($B/B_0$), and the lower panels show ion VDFs averaged in the domain highlighted by the magenta and red boxes. Here, $B_0$ is the background magnetic field strength upstream of the shock front.}
         \label{fig:PIC}
\end{figure*}

For this study, a hybrid PIC simulation was conducted to obtain a more general interpretation of the in situ observations. In Figure ~\ref{fig:PIC} we report results from the simulations performed as discussed in this section. An overview of the magnetic field magnitude is shown in panel (a) of Figure ~\ref{fig:PIC}, showing a fully developed shock transition for the simulation time $T \Omega_{ci} = 100$. The magnetic field associated with the shock front and upstream is highly complex. The shock ramp displays significant perturbations that manifest as an irregular magnetic profile along the tangential direction of the shock front. The intricate shock profile is coupled to the magnetic structures upstream, which effectively has a feedback effect on the shock and feeds further irregular structures as the simulation advances, as studied in detail by \citet{Preisser2020b} and \citet{kajdic2022}. It may be noted that no preexisting upstream turbulence was included in this simulation, which could modify this scenario of the growth of upstream structures and convection to the shock front ~\citep{Trotta2021,Trotta2022a}. 

The consequence of this shock structuring is that depending on how a spacecraft encounters such a shock, the local $\theta_{bn}$ may differ substantially from the global shock normal. To assess the impact on the ion dynamics of the shock structuring, we studied ion VDFs both near the shock and farther upstream.

The bottom panels of Figure~\ref{fig:PIC} display distributions that are averaged over the area in their respective boxes in the top panel, characterizing the ion VDFs upstream (panel (b), magenta box) and near the shock ramp (panel (c), red box). Both distributions prominently feature incoming ions, which are recognizable by their small narrow core. Interestingly, there is a second ion population, suggesting that the shock reflects a component of the incident ions. The second ion population manifests as a diffuse crescent feature that becomes stronger toward the shock front. The spread-out nature of the distribution is somewhat due to the spatial average being analogous to the Solar Orbiter distribution, conveying that this distribution of ions arrived from a highly perturbed shock front. This feature closely resembles the features measured by Solar Orbiter and plotted in Figure \ref{fig:shock_vdfs}. Therefore, the simulations also demonstrate evident signatures of backstreaming ions that are modified by the shock surface perturbations.

\section{Discussion}
\label{sec:discusssion}
Studies that explicitly focus on ion reflection at IP shocks are rare because the typically observed Mach numbers are low. However, our survey of IP shocks since the start of the mission found that Solar Orbiter was fortunate to observe a high Alfv\'en Mach number IP shock on 30 October 2021 with unambiguous evidence of reflected ions. Interestingly, there is evidence that the backstreaming ions may be modulated by nonplanary features of the shock, which was supported by test-particle analysis and hybrid PIC modeling. In addition, whistler waves were found to be likely generated by ion reflection. This study sheds additional light on the connection between reflected ions and the complex shock structure. These results are discussed in more detail below and are placed into context with existing studies.

\subsection{Evidence of shock irregularities}
%% Ripling, non-planarity, large amplitude whistler wave
Ion reflection is affected by shock irregularities, and in this study, both were observed. The shock examined in this study displayed clear signs of a substructure at the shock ramp that was oblique to the upstream magnetic field and the shock normal direction ($\sim 25^\circ$). Comparable features were reported by \citet{kajdic2019}, who analyzed irregularities at IP shocks and concluded that locally, the shock normal can vary substantially from the global or average normal direction. This is important because when an IP shock is measured in situ, parameters such as $\theta_{bn}$ and $M_A$ may not resemble the larger-scale shock structure. In addition, \citet{kajdic2022} explored this further using local simulations and proposed that this may also be linked to the transmission of ULF waves across the shock front. Nonplanar changes in the magnetic field accompanied by density changes have been observed at shocks on numerous occasions. For example, \citet{dimmock2022} reported similar structures at the Venus bow shock during a Solar Orbiter flyby. The characteristics described above have been physically described by some as conditions of shock nonstationarity, which has been attributed to multiple shock processes such as rippling \citep{lowe2003,johlander2016,johlander2018}, wave breaking (also known as gradient catastrophe) \citep{krasnoselskikh2002,dimmock2019}, and corrugation instabilities.

Shock rippling is comparable to a surface wave propagating along the shock that is tangential to the shock normal direction. As a spacecraft transits the shock front, it can cross one or several ripples, constructing a more complex shock encounter with deviations in the magnetic field, density, and modulations in the ion VDF, as shown by \citet{johlander2016,johlander2018} using MMS data for the supercritical terrestrial bow shock. Although the parameters of the \citet{johlander2016} event were similar to the current case, they indicated that the ripples were linearly polarized, whereas the structure we observed was not. A more plausible candidate is large-amplitude whistler waves such as those analyzed by \citet{wilson2012,wilson2017}, which can reach amplitudes equal to or larger than the shock compression jump. The $\theta_{kn}\sim 24^\circ$ is reasonable because \citet{wilson2017} reported similar values for some events. The circular polarization of the structure in this shock is also consistent with the magnetosonic whistler wave that can be caused by shock macrodynamics or instabilities. \citet{dimmock2019} and \citet{dimmock2022} also observed shock structures inside the shock ramp for the Earth and Venus bow shocks, respectively, with several consistent characteristics. Nevertheless, one caveat is that the Mach number of this shock implies that we are unlikely to be in a nonstationary regime according to the gradient catastrophe mechanism \citep{krasnoselskikh2002,dimmock2019} because the nonlinear critical Mach number $M_{nw} = |\cos{\theta_{bn}}|/\sqrt{2m_e/m_p}$ becomes large for oblique shocks, over 20 in this case, when $\theta_{bn}=44^\circ$. In addition, we would expect a smaller $\theta_{kn}$ for the gradient catastrophe mechanism compared to the value that was calculated here.

\subsection{Modulation of backstreaming ions by shock irregularities}
As presented above, this analysis suggests that shock irregularities may play a role in modulating the reflected ion beam and causing some of the complex features revealed in the ion VDFs upstream of the shock. Evidence for this was found in both the Solar Orbiter observations and in supporting test-particle and model results.

\subsubsection{Evidence from Solar Orbiter}
% Field-aligned beam and its modulation by the shock structures
The ion VDFs observed by Solar Orbiter were relatively spread out in phase space, which according to the test-particle analysis and simulation could be due to shock-front irregularities. The analysis of the ion VDFs (see Figure \ref{fig:shock_vdfs}) at varying distances from the shock offered compelling evidence that field-aligned ion beams were observable several minutes upstream of the shock front. Similar features have been reported in previous studies at IP shocks \citep{kajdic2017}, where they typically appear as small anisotropic features with the peak along the background magnetic field direction \citep{paschmann1981}. However, in the case of this shock, the feature is substantially spread across velocity space and is also highly structured. One interpretation is that backstreaming particles are observed from numerous points on the shock due to the distorted nature of the shock front. This could lead to a broadened field-aligned beam feature and to the appearance of multiple beams, as was observed for this shock.

Another explanation for the ion VDF features could be intermediate distributions, which have been shown in previous studies \citep{paschmann1981, fuselier1986, kajdic2017} to exhibit similar characteristics. The field-aligned beam drives a right-hand resonant beam instability that drives in magnetohydrodynamics (MHD)-type waves that disrupt the field-aligned beam, which result in intermediate ions \citep[see ][ and references therin]{fuselier1986}. This may eventually also lead to diffuse ion distributions. Therefore, waves like this would be expected to be observed in concert with intermediate ion distributions. One argument against this mechanism is that we do not see any evidence of MHD waves upstream of the shock (see Figure \ref{fig:shock_overview}k). Still, this is not conclusive because \citet{fuselier1986} reported that they did not see waves like this for half of their events when intermediate ion populations were present. Furthermore, the spread that we observe is highly structured (multiple peaks and separate beam features) and not a clear crescent like a kidney, with a single peak. This may suggest the possibility of multiple beams rather than one single population. Thus, our interpretation is that we observed field-aligned beam(s) that are modulated by the irregular structure of the shock front.

\subsubsection{Further evidence provided by simulations}
The simulations shown in Figure \ref{fig:PIC} provide an insight into how these broad features may be created from a distorted shock front. The VDFs shown in the lower panels are spatially averaged over the respective colored boxes. They show a dispersed field-aligned beam feature that is created by numerous field-aligned beams detected from the perturbed shock front, that is, arriving from different parts of the shock. Similarly, the test-particle analysis provided additional evidence that the features of backstreaming ions are more dispersed for a nonplanar shock, but did not seem to affect the overall fraction of ions that were reflected. However, care should be taken in a comparison with the observations because in that situation, we analyze a local measurement and the broadening may be due to other factors, such as the acquisition time of the VDF. Additionally, simultaneous beams observed from different parts of the shock reaching the spacecraft location may also play a role. Nevertheless, a targeted study focused on model-data comparisons at rippled shocks is required.

The shock distortions may also not only develop from shock ripples or nonstationarity, but from a complex and spatially structured solar wind ahead of the shock front. Evidence can be found in Figure \ref{fig:shock_overview}, revealing that $\mathbf{B_u}$ is not stable, but experiences some rotations and depressions farther upstream. This could imply that along the shock surface, the upstream conditions deviate locally, resulting in an elaborate shock-front profile that is expected to affect the reflection of ions. Thus, the field-aligned beam that Solar Orbiter observes could be modulated by this behavior, effectively manifesting as a spread in velocity space in the VDF. Further evidence of this is that the Mach number of the same shock observed at Wind was considerably lower, signifying that the global shock properties may not be compatible with those calculated by the in situ Solar Orbiter observations.

\subsection{Link between observed whistler waves and ion reflection}
Waves at multiple frequencies upstream of collisionless shocks are known to be linked to reflected ions. In this study, waves with a frequency of approximately 2.5 Hz in the spacecraft frame were observed upstream of the shock near the foot region. Previous research has explored numerous sources of these waves, including proton-beam instabilities \citep{wong1988}, temperature anisotropies \citep{hull2012}, kinetic cross-field streaming instabilities \citep{wu1984,lalti2022}, and shock macrodynamics \citep{krasnoselskikh1985}. The properties of these waves (e.g., $f_{prf}$, $\theta_{kb}$, and $\theta_{kn}$) are comparative to the whistler waves studied by \citet{lalti2022}, and they are also observed in a similar location within the shock foot. Lalti et al. discovered that the kinetic cross-field instability was the most likely generation mechanism, where the reflected ions interact with the incident solar wind electrons. According to \citet{wu1984}, this is expected to occur for oblique whistlers for $\theta_{kb}$ between 30$^\circ$ and 60$^\circ$ for 0.5-1$\omega_{lh}$, which is consistent with the findings in this study. 
Our interpretation is that the kinetic cross-field streaming instability is the likely candidate due to the gyrating ions, which is a direct connection between the waves described above and ion reflection.

Nearer to the shock front, there is some indication of multiple ion populations that are not as closely aligned with $\mathbf{B_u}$ as those further upstream. This could be evidence of gyrating ions, similar to those reported by \citet{cohen2019}, which would be expected from a supercritical quasi-perpendicular shock. A caveat, however, is that the available cadence of the ion VDFs may not be sufficient to resolve this feature. This is because IP shocks have high speeds (compared to the Earth's bow shock) and gyrating ions occur locally within the shock foot, which is typically only measured for a few seconds. Nevertheless, there is some evidence from the presence of whistlers that both gyrating and backstreaming ions may be present for this event.

\subsection{Particle energies}
Particle acceleration at shocks is a fundamental plasma process that is frequently related to particle reflection, and it has received significant attention \citep[e.g.,][]{perri2022}. Therefore, it is worthwhile to consider the energies of shock-related particles for this event. Previous observations of IP shocks have shown populations of energetic particles. An example is the $<200$ keV diffuse ions reported by \citet{kajdic2017}; however, these were not observed for this event. Instead, ion energies reached moderate suprathermal levels of about 15 keV upstream, but dropped suddenly after the ramp. The energies of the observed field-aligned beams were similar to those at the Earth's bow shock, about 10 keV. The argument has been proposed by \citet{kajdic2017} and other authors \citep[e.g.,][]{paschmann1981} that the duration of field-line connection of IP shocks could create higher-energy field-aligned beams than are observed at planetary shocks, where the connection time is shorter. We found no evidence of strong shock-acceleration processes for this event. Our conclusion is that ion reflection and the associated acceleration were similar to the terrestrial bow shock under comparable conditions. This is supported by the similar ion VDF features and energies between this shock and those reported at the terrestrial bow shock. For further comparison, we computed the acceleration efficiency for this shock and others in our shock survey when the calculations were feasible and found values comparable to those obtained statistically for the Earth's bow shock. This further supports the similarities. These calculations are provided in Appendix \ref{appendix:acc_eff}. Although shock irregularities such as large-amplitude waves, ripples, nonstationarity, and other complex structures such as shocklets \citep{wilson2009,trotta2023} may play a role in particle acceleration, the evidence found here is that they can act to modulate reflected ion beams and also result in irregular ion injection; the latter is the focus of a separate study using EPD data for this event.

\section{Conclusions}
The main conclusions from this study are first, that a survey of Solar Orbiter shocks was conducted until the end of August 2022 and the results are provided in Table \ref{tab:shock_list}. Second, most shocks have $M_A$ up to 3 and do not exhibit clear ion reflection, except for a shock on 30 October 2021. Third, a local depression in the magnetic field played a large role in the high $M_A$ for this event as shown in Figure \ref{fig:shock_icme}. Fourth, measurements reveal backstreaming ions up to several minutes upstream of the shock ramp that are field-aligned. Fifth, analysis of the magnetic shock structure suggests that irregularities (ripples, whistler waves, and corrugations) may play a strong role in modulating the features of backstreaming ions. Sixth, the Solar Orbiter observations presented in this study are in strong agreement with test-particle analysis and hybrid PIC modeling. Seventh, whistler waves were consistent with generation via the kinetic cross-field streaming instability, and therefore are directly related to ion reflection. The observations presented here shed light on how the dynamics of ion reflection at IP shocks may be affected by complex shock structures. In addition, this reiterates the difficulty of interpreting localized observations of interplanetary shocks as they may not be representative of global behavior.

\begin{acknowledgements}
APD received financial support from the Swedish National Space Agency (Grant \#2020-00111). MG was supported by the European Unions Horizon 2020 research and innovation program under grant agreement No. 101004131 (SHARP). Work on EPD-STEP was partially funded by the German Space Agency (Deutsches Zentrum für Luft- und Raumfahrt, e.V., (DLR)), grant number 50OT2002. DT has received funding from the European Unions Horizon 2020 research and innovation programme under grant agreement No. 101004159 (SERPENTINE, www.serpentine-h2020.eu). The Solar Orbiter Solar Wind Analyser (SWA) data are derived from scientific sensors which have been designed and created, and are operated under funding provided in numerous contracts from the UK Space Agency (UKSA), the UK Science and Technology Facilities Council (STFC), the Agenzia Spaziale Italiana (ASI), the Centre National d’Etudes Spatiales (CNES, France), the Centre National de la Recherche Scientifique (CNRS, France), the Czech contribution to the ESA PRODEX programme and NASA. Solar Orbiter SWA work at UCL/MSSL is currently funded under STFC grants ST/T001356/1 and ST/S000240/1. XBC thanks PAPIIT DGAPA grant IN110921.
\end{acknowledgements}


   \bibliographystyle{aa} % style aa.bst
  % \bibliography{references} % your references Yourfile.bib

\clearpage
\begin{appendix}
\onecolumn
\section{List of Solar Orbiter interplanetary shocks}
The results of our Solar Orbiter IP shock survey are listed in Table \ref{tab:shock_list}.

\label{appendix:shock list}
\begin{table*}[htb]
\caption{Interplanetary shocks measured by Solar Orbiter. \label{tab:shock_list}}
\centering
\begin{tabular}{l c c c c c c c c c}
\hline
\# & Date  & Time (UTC) & $|R|$ [AU] & $\theta_{bn}$ [$^\circ$] & $M_A$ & $M_A (proxy)$ & $|\mathbf{B_{us}}|/|\mathbf{B_{ds}}|$ & $N_{us}/N_{ds}$ & $V_{sh}$ [km/s]\\
\hline
1$^a$ & 2020-04-19 & 05:06:18 & 0.8 & 47 &  & 2.2 & 1.9 &  & \\
2$^a$ & 2020-08-21 & 19:17:06 & 0.9 & 72 &  & 1.6 & 1.7 &  & \\
3 & 2020-09-17 & 11:44:17 & 1.0 &  &  &  &  & 1.5 & \\
4$^a$ & 2020-11-12 & 23:26:57 & 0.9 & 72 &  & 1.7 & 1.6 &  & \\
5$^a$ & 2020-11-14 & 19:16:14 & 0.9 & 80 &  & 1.2 & 1.4 &  & \\
6$^a$ & 2020-12-06 & 15:14:25 & 0.8 & 58 &  & 1.4 & 1.5 &  & \\
7$^a$ & 2020-12-14 & 02:49:49 & 0.8 & 78 &  & 0.9 & 1.3 &  & \\
8$^a$ & 2021-04-15 & 20:22:46 & 0.8 & 12 &  & 2.0 & 1.8 &  & \\
9$^a$ & 2021-06-07 & 20:02:24 & 1.0 & 29 &  & 2.2 & 1.7 &  & \\
10 & 2021-06-13 & 10:08:41 & 0.9 & 74 & 1.5 & 1.2 & 1.4 & 1.5 & 441\\
11$^a$ & 2021-06-23 & 22:15:31 & 0.9 & 73 &  & 1.7 & 1.7 &  & \\
12$^a$ & 2021-06-27 & 05:54:53 & 0.9 & 17 &  & 2.4 & 1.4 &  & \\
13 & 2021-07-18 & 17:57:54 & 0.8 & 83 & 2.8 & 1.7 & 1.6 & 1.5 & 334\\
14 & 2021-07-19 & 08:28:02 & 0.8 & 67 & 2.8 & 1.7 & 1.7 & 1.7 & 372\\
15 & 2021-07-31 & 00:39:37 & 0.8 & 63 & 1.7 & 1.6 & 1.6 & 2.2 & 380\\
16 & 2021-09-25 & 18:26:07 & 0.6 &  &  &  &  & 3.4 & \\
17 & 2021-10-11 & 07:32:24 & 0.7 & 66 & 2.6 & 1.6 & 1.6 & 1.7 & 469\\
18 & 2021-10-14 & 23:13:06 & 0.7 & 51 & 1.1 & 1.5 & 1.4 & 2.5 & 214\\
19 & 2021-10-30 & 22:02:09 & 0.8 & 44 & 6.7 & 7.4 & 3.6 & 3.5 & 348\\
20 & 2021-11-03 & 12:28:04 & 0.8 & 34 & 2.8 & 1.3 & 1.4 & 1.3 & 476\\
21 & 2021-11-03 & 14:04:26 & 0.8 & 37 & 5.1 & 2.9 & 2.6 & 1.6 & 577\\
22 & 2021-11-16 & 04:01:35 & 0.9 & 66 & 3.4 & 1.3 & 1.4 & 1.3 & 178\\
23 & 2021-11-27 & 22:59:45 & 1.0 & 72 & 2.8 & 2.6 & 2.3 & 2.9 & 391\\
24 & 2021-12-27 & 10:16:33 & 1.0 & 79 & 1.8 & 1.6 & 1.6 & 2.1 & 285\\
25 & 2022-01-08 & 01:51:08 & 1.0 & 82 & 3.8 & 1.5 & 1.5 & 1.9 & 284\\
26 & 2022-02-16 & 21:44:55 & 0.7 & 45 & 1.8 & 1.0 & 1.3 & 1.2 & 480\\
27 & 2022-02-21 & 14:32:20 & 0.7 & 65 & 2.8 & 1.5 & 1.5 & 1.4 & 570\\
28 & 2022-03-08 & 14:45:59 & 0.5 & 58 & 1.1 & 1.5 & 1.5 & 2.0 & 367\\
29 & 2022-03-08 & 21:32:56 & 0.5 & 69 & 3.7 & 2.4 & 2.0 & 2.0 & 366\\
30 & 2022-03-11 & 19:52:14 & 0.4 & 21 & 3.1 & 2.7 & 2.3 & 3.1 & 703\\
31 & 2022-04-03 & 04:51:33 & 0.4 & 47 & 2.7 & 1.7 & 1.7 & 2.1 & 790\\
32 & 2022-04-08 & 13:48:53 & 0.4 & 44 & 7.6 & 2.7 & 2.5 & 1.5 & 545\\
33 & 2022-04-14 & 08:51:56 & 0.5 & 21 & 0.8 & 1.3 & 1.2 & 1.5 & 661\\
34 & 2022-05-08 & 08:15:14 & 0.8 & 49 & 2.9 & 1.9 & 1.7 & 1.8 & 391\\
35 & 2022-05-21 & 14:51:12 & 0.9 & 59 & 4.1 & 2.5 & 2.2 & 2.5 & 339\\
36$^a$ & 2022-06-08 & 12:04:27 & 1.0 & 42 &  & 2.3 & 2.1 &  & \\
37 & 2022-06-10 & 22:55:40 & 1.0 & 22 & 2.6 & 1.7 & 1.6 & 1.8 & 663\\
38 & 2022-06-17 & 00:40:07 & 1.0 & 16 & 9.6 & 5.5 & 2.2 & 2.9 & 385\\
39 & 2022-06-28 & 08:09:09 & 1.0 & 46 & 5.3 & 3.4 & 2.5 & 3.0 & 647\\
40 & 2022-07-03 & 06:00:04 & 1.0 & 50 & 4.2 & 2.7 & 2.3 & 2.1 & 482\\
41 & 2022-07-21 & 09:29:52 & 1.0 & 78 & 2.6 & 1.3 & 1.5 & 1.4 & 440\\
42 & 2022-07-25 & 06:22:48 & 1.0 & 65 & 9.7 & 4.1 & 3.1 & 2.1 & 870\\
43 & 2022-08-01 & 14:17:52 & 1.0 & 87 & 1.4 & 1.2 & 1.4 & 1.4 & 420\\
44 & 2022-08-18 & 02:56:45 & 0.9 & 80 & 2.5 & 1.6 & 1.6 & 1.7 & 218\\
45 & 2022-08-29 & 11:06:39 & 0.8 & 47 & 3.7 & 1.9 & 1.5 & 1.6 & 285\\
46 & 2022-08-30 & 13:02:05 & 0.8 & 81 & 3.2 & 2.3 & 2.1 & 2.1 & 610\\
47 & 2022-08-31 & 21:44:28 & 0.7 & 36 & 6.0 & 3.3 & 2.4 & 2.8 & 1186\\
\hline
\multicolumn{9}{l}{$^{a}$Magnetic coplanarity applied for $\theta_{bn}$}
\end{tabular}
\end{table*}

\clearpage
\section{Calculation of the shock parameters}
\label{app:shock_cal}
Any analysis of the shock structure and particle distributions would be incomplete without determining the shock Mach number ($M_A$) and the angle ($\theta_{bn}$) between the shock normal and the upstream magnetic field. To calculate $M_A$, it is necessary to determine the shock rest frame, which in turn requires knowledge of the shock speed ($V_{sh}$). In this study, the shock speed was calculated according to ion mass flux conservation as follows:
\begin{equation}
    V_{sh} = \frac{(N_d\mathbf{V}_d - N_u\mathbf{V}_u) \cdot \hat{\mathbf{n}}}{N_d-N_u}.
    \label{eq:vsh}
\end{equation}
The upstream velocity in the normal incidence frame (where the shock is at rest) can then be determined from
\begin{equation}
    \mathbf{V}_u' = \left(\mathbf{V}_u\cdot\hat{\mathbf{n}} - V_{sh}\right)\hat{\mathbf{n}}.
\end{equation}
Thus, the Alfv\'en Mach number can be computed from $M_A = |\mathbf{V}_u' \cdot \hat{\mathbf{n}}| / v_A$, where $v_A$ is the Alfv\'en speed ($v_A = B_u/\sqrt{\mu_0 N_u m_p}$).
Using only the magnetic field measurements, $M_A$ can also be estimated using the proxy devised by \citet{gedalin2021},
\begin{equation}
    \frac{B_{m}}{B_u} = \sqrt{2 M^2 \left(1 - \sqrt{1-s} \right)+1}. 
    \label{proxy}
\end{equation}
In equation \ref{proxy}, $B_m$ is the maximum magnetic field across the shock, $s$ is the normalized potential jump $s = 2\phi_{NIF}/m_p V_u^2$, and $\phi_{NIF}$ is the cross-shock electrostatic potential. To calculate $M_A$, equation \ref{proxy} was solved for $M$ based on $s=0.6$.

The fast magnetosonic Mach number is related to the Alfv\'en Mach number by $(M=M_A(v_A/v_f)$), where 
\begin{eqnarray}
&& v_A^2=\frac{B_u^2}{\mu_0 N_um_p}\\
&& v_s^2=\frac{\gamma (T_e+T_p)}{m_p}\\
&& v_f^2 =\frac{1}{2} \left[(v_A^2+v_s^2)+\sqrt{(v_A^2+v_s^2)^2-4v_A^2v_s^2\cos^2\theta_{bn}}\right].
\end{eqnarray}
Here, \(m_p\) is the proton mass, and it is where the upstream proton and electron temperatures are equal, \(T_p=T_e\), while the adiabatic index \(\gamma=5/3\). 

Obtaining a reliable shock normal is not always straightforward \citep{paschmann1998}. Therefore, several methods were applied, including mixed-mode coplanarity, magnetic coplanarity, and minimum variance. These methods agreed relatively well, but the existence of waves and magnetic structures near the shock meant that the methods that relied solely on the magnetic field were less precise. Therefore, the mixed-mode coplanarity normal direction was used for all shock calculations. The projection of the magnetic field along the shock normal was also visually checked to ensure the lack of any significant offset from upstream-downstream, although the shock structure did result in some perturbations at the ramp.

\section{Test particle analysis setup} \label{appendix: test_particle}
\subsection{Planar shock}
The shock magnetic field is set up as follows:
\begin{eqnarray}
    \label{shock_planar}
    &B_x& = \cos{\theta_{bn}}  \nonumber\\
    &B_y& = A_2 \sin{\theta_{bn}} S(x) \sin{\phi(x)}  \nonumber \\ 
    &B_z& = \sin{\theta_{bn}}(1 + (A_1 + A_2 \cos{\phi(x)}))S(x))
,\end{eqnarray}
where
\begin{eqnarray}
    \label{shock_planar}
    &S(x)& = \frac{1}{2}\left(  1 + \tanh{\frac{3x}{D}} \right) \nonumber\\
    &\phi(x)& = 2 \pi S(x)  \nonumber.
\end{eqnarray}
The electric field is defined as
\begin{eqnarray}
    \label{shock_planar}
    &E_x& = \frac{S_{NIF}}{2(B_d-1)} \left(\frac{d |\mathbf{B}|}{dx}\right)  \nonumber\\
    &E_y& =\sin{\theta_{bn}}  \nonumber \\ 
    &E_z& = 0
,\end{eqnarray}
where $B_d = \sqrt{(1+A_1+A_2)^2 \sin^2{\theta_{bn}} +  \cos^2{\theta_{bn}}}$ and $S_{NIF}$ is the normalized cross-shock potential
\begin{equation}
    S_{NIF} = \frac{2e\phi_{NIF}}{m_p {V_u}^2}.
\end{equation}

\subsection{Rippled shock}
Strong variations of $B_n$ inside the ramp indicate spatial inhomogeneity along the shock front or time dependence. We included modeling of rippling as follows. We started by representing a stationary planar profile using the vector and scalar potentials,
\begin{equation}
    B_x = -\frac{\partial A_y}{\partial z}
\end{equation}
\begin{equation}
    B_y = -\frac{\partial A_z}{\partial x}
\end{equation}
\begin{equation}
    B_z = \frac{\partial A_y}{\partial x}
\end{equation}
\begin{equation}
    E_x = -\frac{\partial \phi}{\partial x}
\end{equation}
\begin{equation}
    E_y = -\frac{\partial A_y}{\partial t}
\end{equation}
\begin{equation}
    E_z = 0
.\end{equation}
Let
\begin{equation}
    A_x = 0
\end{equation}
\begin{equation}
    A_y = A_y(x)-B_{x0}z - E_{yo}t
\end{equation}
\begin{equation}
    A_z = A_z(x)
\end{equation}
\begin{equation}
    \phi = \phi(x)
.\end{equation}
Now let
\begin{equation}
    A_x = 0
\end{equation}
\begin{equation}
    A_y = A_y(X)-B_{x0}z -E_{y0}t
\end{equation}
\begin{equation}
    A_z = A_z(X)
\end{equation}
\begin{equation}
    \phi = \phi(X)
\end{equation}
\begin{equation}
    X = x + \psi(x,y,z) = x + f(y,z,t)g(x)
,\end{equation}
so that
\begin{equation}
    B_x = \frac{\partial A_z}{\partial y} - \frac{\partial A_y}{\partial z} = \frac{\partial A_z}{\partial X}X_y - \frac{\partial A_y}{\partial X}X_z + B_{x0} = -B_y(X)X_y + B_z(X)X_z + B_{x0}
\end{equation}
\begin{equation}
B_y = -\frac{\partial A_z}{\partial x}=-\frac{\partial A_z}{\partial X}X_x = B_y(X)X_x
\end{equation}
\begin{equation}
B_z = \frac{\partial A_y}{\partial x} = \frac{\partial A_y}{\partial X}X_x = B_z(X)X_x
\end{equation}
\begin{equation}
E_x = \frac{\partial \phi}{\partial X}X_x = E_x(X)X_x
\end{equation}
\begin{equation}
E_y = -\frac{\partial \phi}{\partial X}X_y - \frac{\partial A_y}{\partial X}X_t + E_{y0} = E_x(X)X_y - B_z(X)X_t = E_{y0}
\end{equation}
\begin{equation}
E_z = -\frac{\partial \phi}{\partial X}X_z - \frac{\partial A_z}{\partial X}X_t = E_x(X)X_z + B_y(X)X_t.
\end{equation}
Eventually,
\begin{eqnarray}
    \label{shock_planar}
    &B_x& = B\cos{\theta_{bn}} - B_y(X)X_y + B_z(X)X_z \nonumber\\
    &B_y& = B_y(X)X_x \nonumber \\ 
    &B_z& = B_z(X)X_x \nonumber \\
    &E_x& = E_x(X)X_z \nonumber\\
    &E_y& = V_uB_u\sin{\theta_{bn}} + E_x(X)X_y - B_z(X)X_t \nonumber \\ 
    &E_z& = E_x(X)X_z + B_y(X)X_t
\end{eqnarray}
For the current analysis, we selected
\begin{equation}
    f(y,z,t) = \sin{k_y + k_zz + \omega t}
\end{equation}
\begin{equation}
    k_y = 0;
\end{equation}
\begin{equation}
    k_z = 2 \pi
\end{equation}
\begin{equation}
    \omega = 2 \pi
.\end{equation}

\begin{figure}[h]
   \centering
   \includegraphics[width=6cm]{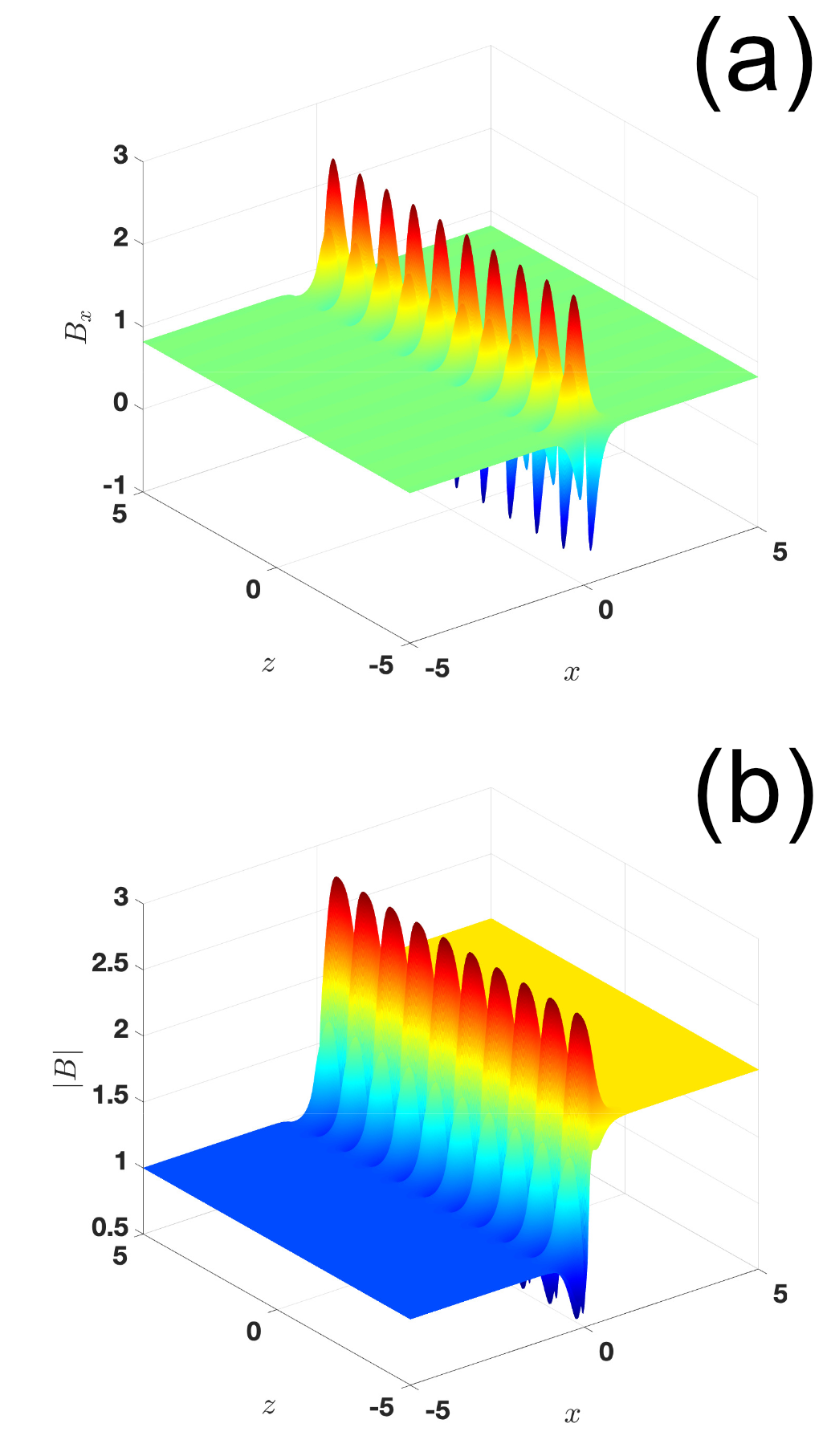}
      \caption{Normal component (a) with model rippling and the magnetic field magnitude (b).}
         \label{fig:test_particle_rippled_shock}
\end{figure}

We plot in Figure \ref{fig:test_particle_rippled_shock} the normal component of the magnetic field and its magnitude for the rippled shock.

\section{Acceleration efficiency}
\label{appendix:acc_eff}
For the Earth's bow shock environment, \citet{johlander2021} and \citet{lalti2022b} used a set of shocks to study energetic ions at the shock front statistically in terms of their acceleration efficiency. We can perform the same task for several of the shocks listed in Table \ref{tab:shock_list}. The acceleration efficiency is defined as
\begin{equation}
    \epsilon(E_0) = \left\langle \frac{U_i(E_i>E_0)}{U_i(E_i>0)} \right\rangle.
    \label{eq:acceleration_efficiency}
\end{equation}
In Equation \ref{eq:acceleration_efficiency}, $U_i(E_i>E_0)$ is the ion energy density downstream of the shock in the local plasma frame that exceeds the threshold energy $E_0$, based on
\begin{equation}
    U_i(E_i>E_0) = 4\pi\sqrt{\frac{2}{m^2_i}}\int^{E_{max}}_{E_0}dE_i\sqrt{E^3_i} f_i(E_i),
\end{equation}
where the threshold $E_0$ is set as ten times the upstream solar wind energy in the downstream frame. Due to the low energies typically measured by SWA-PAS at IP shocks, it was not feasible to compute $\epsilon$ for all the shocks listed in Table \ref{tab:shock_list}, but we computed it for a total of seven of them, which are listed in Table \ref{tab:epsilon}.
\begin{table}
\caption{Values of acceleration efficiency ($\epsilon$) as well as the shock parameters for the shocks that could be evaluated.}
\label{tab:epsilon}
\centering
\begin{tabular}{l l l l l}
\hline
Time (UTC) & \# & $\epsilon$ (\%)  & $M_A$ & $\theta_{bn}$  \\
\hline
2021-10-30 22:02  & 19 & 3.4  & 6.7 & 44 \\
2021-11-03 14:04  & 21 & 16.2 & 5.1 & 37 \\
2022-04-08 13:48  & 32 & 0.8  & 7.6 & 44 \\
2022-06-17 00:40  & 38 & 12.3  & 9.6 & 16 \\
2022-06-28 08:09  & 39 & 2.4  & 5.3 & 46 \\
2022-07-25 06:22  & 42 & 0.5  & 9.7 & 65 \\
2022-08-31 21:44  & 47 & $\sim$0  & 6.0 & 36 \\
\hline
\end{tabular}
\end{table}
The values of the acceleration efficiency for shocks listed in Table \ref{tab:epsilon} are generally lower than the value of the bow shock \citep{johlander2021,lalti2022b}, and most values fall between 5\%-15\%. The event studied in detail here has a value of 3.4\%. According to \citet{lalti2022}, this is reasonable, but values between 5\%-10\% would be expected from statistics. Although it should be noted that the general trend is consistent, lower $\theta_{bn}$ shocks were more efficient at accelerating ions. A caveat here is that these calculations could not be performed for the majority of shocks, and we did not include EPD data. However, EPD data were checked (see Figure \ref{fig:shock_overview}g) for the shock on 30 October, and there was no significant energetic ion population downstream. As a result, we interpret the value of $\epsilon$ as a reasonable estimate for this shock because EPD observations should not affect this value.

\end{appendix} 

\end{document}